\documentclass{aa}

\usepackage{graphicx}
\usepackage{txfonts}
\begin{document}

   \title{The GAPS programme with HARPS-N at TNG}

   \subtitle{XVI. Measurement of the Rossiter-McLaughlin effect of  transiting planetary systems HAT-P-3, HAT-P-12, HAT-P-22, WASP-39, and WASP-60}

   \author{
          L. Mancini \inst{1,3,4}
          \and
          M. Esposito\inst{2}
          \and
          E. Covino\inst{2}
          \and
          J. Southworth\inst{5}
          \and
          K. Biazzo\inst{6}
          \and
          I. Bruni\inst{7}
          \and
          S. Ciceri\inst{8}
          \and
          D. Evans\inst{5}
          \and
          A.~F. Lanza\inst{6}
          \and
          E. Poretti\inst{9}
          \and
          P. Sarkis\inst{3}
          \and
          A.~M.~S. Smith\inst{10}
          \and
          M. Brogi\inst{11}
          \and
          L. Affer\inst{12}
          \and
          S. Benatti\inst{13}
          \and
          A. Bignamini\inst{14}
          \and
          C. Boccato\inst{13}
          \and
          A.~S. Bonomo\inst{4}
          \and
          F. Borsa\inst{9}
          \and
          I. Carleo\inst{13,15}
          \and
          R. Claudi\inst{13}
          \and
          R. Cosentino\inst{16}
          \and
          M. Damasso\inst{4}
          \and
          S. Desidera\inst{13}
          \and
          P. Giacobbe\inst{4}
          \and
          E. Gonz\'{a}lez-\'{A}lvarez\inst{12}
          \and
          R. Gratton\inst{13}
          \and
          A. Harutyunyan\inst{16}
          \and
          G. Leto\inst{6}
          \and
          A. Maggio\inst{12}
          \and
          L. Malavolta\inst{13,15}
          \and
          J. Maldonado\inst{12}
          \and
          A. Martinez-Fiorenzano\inst{16}
          \and
          S. Masiero\inst{12}
          \and
          G. Micela\inst{12}
          \and
          E. Molinari\inst{16,17}
          \and
          V. Nascimbeni\inst{15,13}
          \and
          I. Pagano\inst{6}
          \and
          M. Pedani\inst{16}
           \and
          G. Piotto\inst{15}
          \and
          M. Rainer\inst{9}
          \and
          G. Scandariato\inst{6}
          \and
          R. Smareglia\inst{15}
          \and
          A. Sozzetti\inst{4}
          \and
          G. Andreuzzi\inst{16,18}
          \and
          Th. Henning\inst{3}
          }

\institute{
Dipartimento di Fisica, Universit\`a di Roma Tor Vergata, Via della Ricerca Scientifica 1, 00133 -- Roma, Italy \\
\email{lmancini@roma2.infn.it}
\and
INAF -- Osservatorio Astronomico di Capodimonte, via Moiariello, 16, 80131 -- Naples, Italy
\and
Max Planck Institute for Astronomy, K\"{o}nigstuhl 17, 69117 -- Heidelberg, Germany
\and
INAF -- Osservatorio Astrofisico di Torino, via Osservatorio 20, 10025 -- Pino Torinese, Italy
\and
Astrophysics Group, Keele University, Keele ST5 5BG, UK
\and
INAF -- Osservatorio Astrofisico di Catania, via S. Sofia 78, 95123 -- Catania, Italy
\and
INAF -- Osservatorio Astronomico di Bologna, Via Ranzani 1, 40127 -- Bologna, Italy
\and
Department of Astronomy, Stockholm University, AlbaNova University Center, 106 91 Stockholm, Sweden
\and
INAF -- Osservatorio Astronomico di Brera, via E. Bianchi 46, 23807 -- Merate (LC), Italy
\and
Institute of Planetary Research, German Aerospace Center, Rutherfordstrasse 2, 12489 -- Berlin, Germany
\and
Department of Physics, University of Warwick, Coventry CV4 7AL, UK
\and
INAF -- Osservatorio Astronomico di Palermo, Piazza del Parlamento 1, 90134 -- Palermo, Italy
\and
INAF -- Osservatorio Astronomico di Padova, Vicolo dell'Osservatorio 5, 35122 -- Padova, Italy
\and
INAF -- Osservatorio Astronomico di Trieste, via G.~B. Tiepolo 11, 34143 -- Trieste, Italy
\and
Dipartimento di Fisica e Astronomia G. Galilei, Universit\`a di Padova, Vicolo dell'Osservatorio 2, 35122 -- Padova, Italy
\and
INAF -- Fundaci\'{o}n Galileo Galilei, Rambla Jos\'{e} Ana Fernandez P\'{e}rez 7, 38712 -- Bre\~{n}a Baja, Spain
\and
INAF -- Osservatorio Astronomico di Cagliari, Via della Scienza 5, 09047 -- Selargius (CA), Italy
\and
INAF -- Osservatorio Astronomico di Roma, via Frascati 33, 00040 -- Monte Porzio Catone (Roma), Italy
}

   \date{Received ; Accepted }


  \abstract
{}
   {We aim to derive the degree of alignment between planetary orbit and stellar spin angular momentum vectors and look for possible links with other orbital and fundamental physical parameters of the star-planet system. We focus on the characterisation of five transiting planetary systems (HAT-P-3, HAT-P-12, HAT-P-22, WASP-39, and WASP-60) and the determination of their sky-projected planet orbital obliquity through the measurement of the Rossiter-McLaughlin effect.}
   {We used HARPS-N high-precision radial velocity measurements, gathered during transit events, to measure the Rossiter-McLaughlin effect in the target systems and determine the sky-projected angle between the planetary orbital plane and stellar equator. The characterisation of stellar atmospheric parameters was performed by exploiting the HARPS-N spectra, using line equivalent width ratios and spectral synthesis methods. Photometric parameters of the five transiting exoplanets were re-analysed through 17 new light curves, obtained with an array of medium-class telescopes, and other light curves from the literature. Survey-time-series photometric data were analysed for determining the rotation periods of the five stars and their spin inclination.}
{From the analysis of the Rossiter-McLaughlin effect we derived a sky-projected obliquity of $\lambda=21.2^{\circ}\pm 8.7^{\circ}$, $\lambda=-54^{\circ}\,^{+41^{\circ}}_{-13^{\circ}}$, $\lambda=-2.1^{\circ}\pm 3.0^{\circ}$, $\lambda=0^{\circ}\pm 11^{\circ}$, and $\lambda=-129^{\circ} \pm 17^{\circ}$ for HAT-P-3\,b, HAT-P-12\,b, HAT-P-22\,b, WASP-39\,b, and WASP-60\,b, respectively. The latter value indicates that WASP-60\,b is moving on a retrograde orbit. 
The stellar activity of HAT-P-22 indicates a rotation period of $28.7 \pm 0.4$\,days, which allowed us to estimate the true misalignment angle of HAT-P-22\,b, $\psi=24^{\circ}\pm{18^{\circ}}$. The revision of the physical parameters of the five exoplanetary systems returned values that are fully compatible with those existing in the literature. The exception to this is the WASP-60 system, for which, based on higher quality spectroscopic and photometric data, we found a more massive and younger star and a larger and hotter planet.}
   {}

   \keywords{Extrasolar planets -- Stars: late-type, fundamental parameters -- Techniques: radial velocities, photometric -- Stars: individual: HAT-P-3; HAT-P-12; HAT-P-22; WASP-39; WASP-60}

   \maketitle
%

\section{Introduction}
\label{sec:introduction}
The study of the physical and orbital properties of extrasolar planets, in connection with the physical characteristics of their host stars, provides important insight into formation and evolution mechanisms of planetary systems, which  are currently a matter of extensive debated. In particular, the evolution theories of planetary orbits are very difficult to establish on solid ground because there are so many possible architectures of planetary systems and many factors can contribute to modify the dynamic of these systems during their secular life.

The existence of a hot-Jupiter population, i.e. Jupiter-mass planets with orbital periods of only a few days, is a clear indication that inward migration occurred during the process of formation or early evolution for many of these gaseous planets\footnote{Possible scenarios of in situ formation of hot Jupiters were also theorised (e.g. \citealp{
bodenheimer:2000,boley:2016,batygin:2016}).}. Widely accepted scenarios of the migration of giant planets, which are supported by hydrodynamic simulations, involve planet-disc interaction, in which planets are kept on circular orbits with orbital axes parallel to the stellar spin axis (e.g. \citealp{lin:1996,marzari:2009,bitsch:2011}), whereas planets on eccentric and oblique orbits can be the result of planet-planet scattering (e.g. \citealp{rasio:1996,chatterjee:2008,marzari:2014}) or Kozai torque by a distant massive companion (e.g. \citealp{fabrycky:2007}).
Therefore, the orbital obliquity, $\psi$, i.e. the angle between the orbital angular momentum and the spin of the host star, represents an extremely important parameter, as we can use it to probe how planetary systems form and evolve. As all the above-mentioned migration scenarios probably occur (e.g. \citealp{nagasawa:2008,nelson:2017}), we must study a statistically significant sample of exoplanetary systems to quantify the relative importance of the orbital obliquity \citep{schlaufman:2010}.

While $\psi$ is a quantity that is difficult to determine, the measurement of its sky projection, $\lambda$, is commonly achievable for stars hosting transiting exoplanets, mainly through the observation of the Rossiter-McLaughlin (RM) effect. This is an anomalous radial velocity (RV)\ variation that occurs when a planet transits a rotating star and can be accurately measured for relatively bright stars with high-precision RV instruments. 
Precise values of $\lambda$ have now been obtained for about a hundred giant exoplanets\footnote{Data taken from TEPCat: \texttt{http://www.astro.keele.ac.uk/jkt/ tepcat/rossiter.html} \citep{southworth:2011}.}, the majority of which show values of $\lambda$ close to zero similar to the planetary bodies orbiting our Sun, although a considerable fraction (nearly 40\%) show substantial misalignment \citep{albrecht:2012}. Of these, ten are in nearly polar orbits and another ten are in retrograde orbits\footnote{Following \citet{addison:2013}, we have considered near-polar orbits as those with spin-orbit angles between
$(3\pi/8)<\lambda< (5\pi/8)$ or $(-3\pi/8) >\lambda> (-5\pi/8)$ and retrograde orbits for spin-orbit angles between $(5\pi/8) \leq \lambda \leq (11\pi/8)$ or $(-5\pi/8) \geq \lambda \geq (-11\pi/8)$.}. Such extreme spin-orbit misalignments may also be explained through secular mutual close encounters, or Kozai-Lidov oscillations of orbital eccentricity and inclination induced by a distant companion, whose orbit is significantly tilted with respect to the orbit of the inner planet \citep{nagasawa:2008,naoz:2011}.

As the number of the measurements of $\lambda$ increased in recent years, several empirical trends were noted, of which the most debated is that between $\lambda$ and the effective temperature, $T_{\rm eff}$, of parent stars. By plotting these two quantities together, one can see that planetary systems having stars with $T_{\rm eff}\lesssim 6250$\,K have good spin-orbit alignment, whereas the values of $\lambda$ for those with $T_{\rm eff}\gtrsim 6250$\,K are more broadly distributed \citep{winn:2010,albrecht:2012}. The suggestion that $\lambda$ seems to increase as the amount of stellar surface convection decreases can be explained by the fact that tidal dissipation is much less efficient for hot stars than the cold stars because the former have a smaller convective zone than the latter \citep{valsecchi:2014}.
However, the presence of several exceptions (e.g. WASP-8: \citealt{queloz:2010,bourrier:2017}; HAT-P-18: \citealt{esposito:2014}, HATS-14: \citealt{zhou:2015}) suggests that the truthfulness of the $\lambda-T_{\rm eff}$ trend has to be verified more accurately by enlarging the sample, especially by exploring the range of low values of $T_{\rm eff}$. This is a critical point because measurements of the RM effect become more arduous for slow-rotating cool stars and require large-aperture telescopes and spectrographs with high performances  since the amplitude of the RM effect is $\propto k^2 \, v \sin{i_{\star}}$; $k$ is the ratio of the planetary to stellar radii and $v \sin{i_{\star}}$ is the projected rotational velocity of the star.

Within the framework of the long-term observational programme Global Architecture of Planetary Systems (GAPS), which uses the high-resolution spectrograph HARPS-N at the 3.5\,m Telescopio Nazionale Galileo (TNG), we conducted a subprogramme for studying the spin-orbit alignment of a large sample of known exoplanetary systems \citep{covino:2013}. Our project is especially focussed on those hosted by relatively cold stars, but still sufficiently bright ($V<14$\,mag) \citep{mancini:2015,esposito:2017}. Moreover, our programme is supported by photometric follow-up observations with an array of medium-class telescopes. We use these observations to record high-quality light curves of planetary-transit events of the targets in our sample list. The main aim is to refine the whole set of physical parameters of the planetary systems and check for possible stellar activity; starspots can actually play an important role in modelling transit light curves (e.g. \citealp{mancini:2017}).
 In this work, we present new detections of the RM effect for five exoplanetary systems for which measurements of $\lambda$ were not available before now.

The paper is organised as follows. In Sect.~\ref{sec:history} we briefly describe the systems that are the subjects of this study. Both spectroscopic and photometric observations are presented in Sect.~\ref{sec:observation}, together with the description of the corresponding data reduction procedures. The analysis of the photometric light curves is discussed in Sect.~\ref{sec:lc_analysis}. Sect.~\ref{sec:frequency_analysis} is devoted to exploring possible stellar activity by analysing the survey-time-series photometric data.
The stellar atmospheric properties and measurements of the spin-orbit relative orientation of the systems, based on HARPS-N data, are presented in Sect.~\ref{sec:spectra_analysis}. Our refinements of the physical parameters of the systems are reported in Sect.~\ref{sec:physical_parameters}, with a particular attention to WASP-60, for which we found values different from those measured by its discoverers. Finally, in Sect.~\ref{sec:discussion} we summarise and discuss the main results of this study.

\section{Case history}
\label{sec:history}

In this work, we present measurements of the RM effect for five transiting exoplanetary systems. These systems are HAT-P-3, HAT-P-12, HAT-P-22, WASP-39, and WASP-60, each of which is composed of a hot giant planet, with an equilibrium temperature, $T_{\rm eq}$, in the range $960-1320$\,K, and a mid-K- or G-type star, with an effective temperature, $T_{\rm eff}$, in the range $4650-5900$\,K. Their main physical parameters, which were also recalculated in this work (see Sect.~\ref{sec:physical_parameters}), are summarised in the tables reported in Appendix~\ref{appendix_PhysicalParameters}. The eccentricity of the orbits of each of the five exoplanet was fixed to zero, according to the values determined by \citet{bonomo:2017}.

\subsection{HAT-P-3}
\label{sec:hatp3}
The planetary system HAT-P-3 is composed of a $V=11.6$\,mag metal-rich, early-K dwarf star, around which a hot Jupiter (mass $\approx 0.6\,M_{\rm Jup}$ and radius $\approx 0.9\,R_{\rm Jup}$) revolves on a circular orbit, producing transit events every 2.9 days with a depth of 1.5\% \citep{torres:2007}. In recent years, several studies of this system were presented, reporting slightly improved values of the physical \citep{torres:2008,gibson:2010,chan:2011,southworth:2012,torres:2012,eastman:2013,ricci:2017} and orbital parameters \citep{torres:2008,madhusudhan:2009,gibson:2010,nascimbeni:2011,chan:2011,pont:2011,sada:2012,southworth:2012,eastman:2013,sada:2016}. Two occultations of HAT-P-3\,b were measured with the {\it Spitzer} space telescope in the 3.6 and 4.5\,$\mu$m bands, from which it was found that the planet has inefficient heat transfer from its day to night side, but it is not clear if there is a temperature inversion in its atmosphere \citep{todorov:2013}. No information is available about either its atmospheric composition or the obliquity of its orbit.

\subsection{HAT-P-12}
\label{sec:hatp12}
Having a mass of $\approx 0.2\,M_{\rm Jup}$, HAT-P-12\,b can be considered as a sub-Saturn type planet \citep{hartman:2009}. It moves on a circular orbit, transiting its parent star, a relatively metal-poor K4\,V star, with a periodicity of 3.2\,days and lowering its brightness by 2\%. The orbital and physical parameters of this planetary system were revised by several authors \citep{lee:2012,knutson:2014,hinse:2015,mallonn:2015,sada:2016} based on new photometric light curves. No occultations of HAT-P-12\,b were detected so far \citep{todorov:2013}. The low density ($\approx 0.24\,\rho_{\rm Jup}$) of the planet and relatively bright primary ($V=12.8$\,mag) make this system a suitable target for transmission spectroscopy. A near-infrared (NIR) transmission spectrum of the planet was presented by \citet{line:2013}, based on data collected with the Hubble Space Telescope (HST); no water-absorption features were observed, suggesting an atmosphere dominated by high-altitude clouds. This result was confirmed by \citet{mallonn:2015}, who extended the analysis to optical wavelengths via broadband photometric observations with a group of professional class telescopes.

\subsection{HAT-P-22}
\label{sec:hatp22}
The massive and compact hot Jupiter HAT-P-22\,b (mass $\approx 2.1\,M_{\rm Jup}$ and radius $\approx 1.1\,R_{\rm Jup}$) circularly orbits a fairly metal-rich and bright ($V=9.7$\,mag) G5\,V star with a period of $3.2$\,days, producing planetary-transit events with a depth of 1.5\% \citep{bakos:2011}. Further studies of this planetary system, mostly based on new photometric light curves, presented slight refinements of the orbital \citep{knutson:2014,turner:2016} and physical parameters \citep{torres:2012,hinse:2015,sousa:2015,basturk:2015,turner:2016}. Occultation measurements of HAT-P-22\,b with {\it Spitzer} were performed by \citet{kilpatrick:2017}, who concluded that its atmosphere does not experience efficient recirculation.

\subsection{WASP-39}
\label{sec:wasp39}
The discovery of the transiting exoplanet WASP-39\,b was announced by \citet{faedi:2011}. It is a highly inflated Saturn-mass planet (mass $\approx 0.3\,M_{\rm Jup}$ and radius $\approx 1.3\,R_{\rm Jup}$) circularly orbiting a late G-type dwarf star with a period of roughly four\,days. Further observations refined the orbital \citep{ricci:2015,fischer:2016,maciejewski:2016} and physical parameters \citep{maciejewski:2016,nikolov:2016}. Occultation measurements with {\it Spitzer} were also carried out for WASP-39\,b, suggesting a very efficient circulation of energy from the day to the night side \citep{kammer:2015}.  A Rayleigh scattering slope as well as sodium and potassium absorption features were detected by \citet{fischer:2016} using HST transit observations and complementary data from {\it Spitzer}. These findings were confirmed by ground-based, transmission-spectroscopy observations, which were obtained with the Very Large Telescope (VLT) \citep{nikolov:2016}.

\subsection{WASP-60}
\label{sec:wasp60}
The hot Jupiter WASP-60\,b  was discovered by \citet{hebrard:2013}, who measured for this planet a mass of $\approx 0.5\,M_{\rm Jup}$ and a radius of $\approx 0.9\,R_{\rm Jup}$. The authors found that it transits in front of its parent star, a G1\,V star with $M_{\star}\approx 1.1\,M_{\sun}$ and $R_{\star} \approx 1.1\,R_{\sun}$, every $\approx 4.3$\,days, producing shallow transits of 0.6\%. These measurements were based on RV data, photometric data from the SuperWASP survey, and on a single incomplete follow-up light curve. As we see in Sect.~\ref{sec:physical_parameters}, several of these findings are not in agreement with the results presented in this work. According to \citet{bonomo:2017}, the eccentricity of the orbit of WASP-60\,b is compatible with zero, but with an uncertainty larger than 0.05. A new incomplete, photometric light curve of a WASP-60\,b transit was reported by \citet{turner:2017}. No further follow-up works have yet been presented for this exoplanetary system.

\section{Observation and data reduction}
\label{sec:observation}

In this section we present new times-series spectroscopic data of HAT-P-3, HAT-P-12, HAT-P-22, WASP-39, and WASP-60. The spectra were obtained with HARPS-N during their transit events with the specific purpose of measuring the RM effects for each of the five exoplanets. We also present new photometric follow-up observations of HAT-P-3, HAT-P-12, and WASP-60.

\subsection{HARPS-N spectroscopic observations}
\label{sec:HARPS-N_observation}

The spectroscopic observations of the transits were carried out using the HARPS-N (High Accuracy Radial velocity Planet Searcher-North; \citealp{cosentino:2012}) spectrograph at the 3.58\,m TNG on the following nights: 2013/06/10 (HAT-P-3), 2013/10/20 (WASP-60), 2014/04/03 (HAT-P-22), 2015/03/13 and 2015/04/24 (HAT-P-12), and  2015/05/04 (WASP-39). The observations were performed with a simultaneous Thorium-lamp spectrum for the stars with $V < 12$ (HAT-P-3 and HAT-P-22) and with fibre A on target and fibre B on sky for the other three fainter stars. The log of the HARPS-N observations is given in Table~\ref{tab:tng-log}.

The reduction of the spectra was performed using the latest version (3.7) of the HARPS-N data reduction software (DRS) pipeline \citep{cosentino:2014,smareglia:2014}. Radial velocity measurements, with corresponding uncertainties, were computed by cross-correlating each spectrum with a numerical template mask \citep{baranne:1996,pepe:2002,lovis:2007}. In addition to RVs, the DRS provides 1-D wavelength-calibrated spectra, which we used for the determination of the atmospheric parameters of the star (see Sect.~\ref{sec:spectra_analysis}), the Mount Wilson S-index, and  the $\log{R'_{\rm HK}}$ chromospheric activity index for stars with $B-V <1.2$ \citep{lovis:2011}. The RV measurements for our targets are reported in Table~\ref{tab:RV_HAT-P-3}, \ref{tab:RV_HAT-P-12}, \ref{tab:RV_HAT-P-22}, \ref{tab:RV_WASP-39}, and \ref{tab:RV_WASP-60} for HAT-P-3, HAT-P-12, HAT-P-22, WASP-39, and WASP-60, respectively. The typical signal-to-noise ratio (S/N) of the extracted spectra is 31, 15, 61, 22, 26  per pixel at 550\,nm for HAT-P-3, HAT-P-12, HAT-P-22, WASP-39, and WASP-60, respectively. We also checked that the RV measurements are not affected by the Moon. Following the method that was adopted by \citet{esposito:2017}, a correction for Moon light contamination was applied by subtracting the cross-correlation function of fibre B from that of fibre A, and then measuring the stellar RV by means of a Gaussian fit to this difference. The values of the RV measurements obtained from this procedure are very similar to the previous measurements and, practically, within the uncertainties in all the five cases.

\begin{figure}
\centering
\includegraphics[width=\hsize]{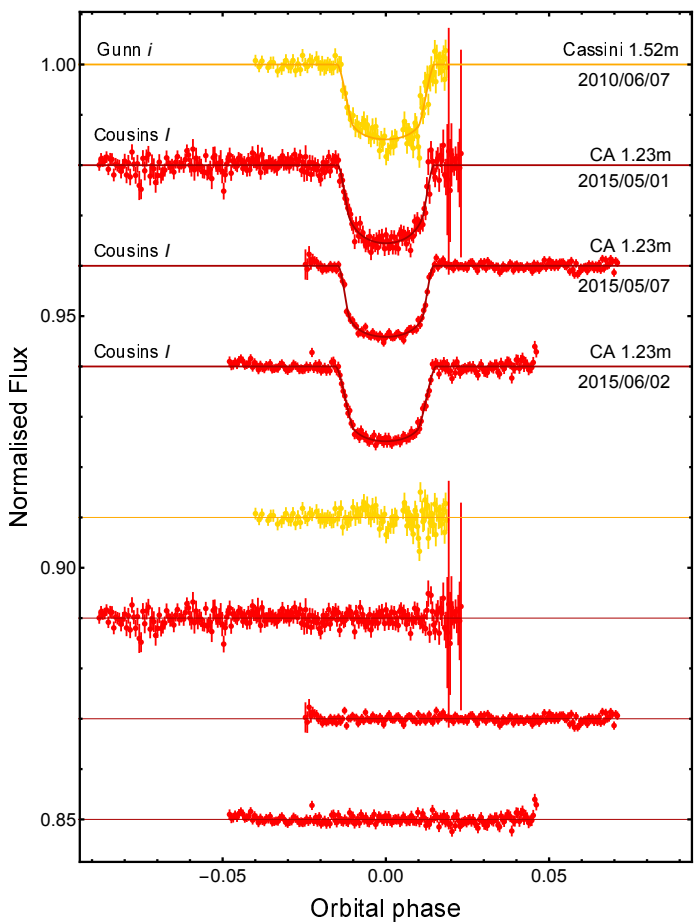}
\caption{Phased light curves of HAT-P-3\,b transits presented in this work. These phased light curves are compared with the best {\sc jktebop} fits. The dates, telescopes, and filters related to the observation of each transit event are indicated. Residuals from the fits are plotted at the base of the figure.}
\label{fig:hatp3_lc}
\end{figure}
%
\begin{figure*}
\centering
\includegraphics[width=18cm]{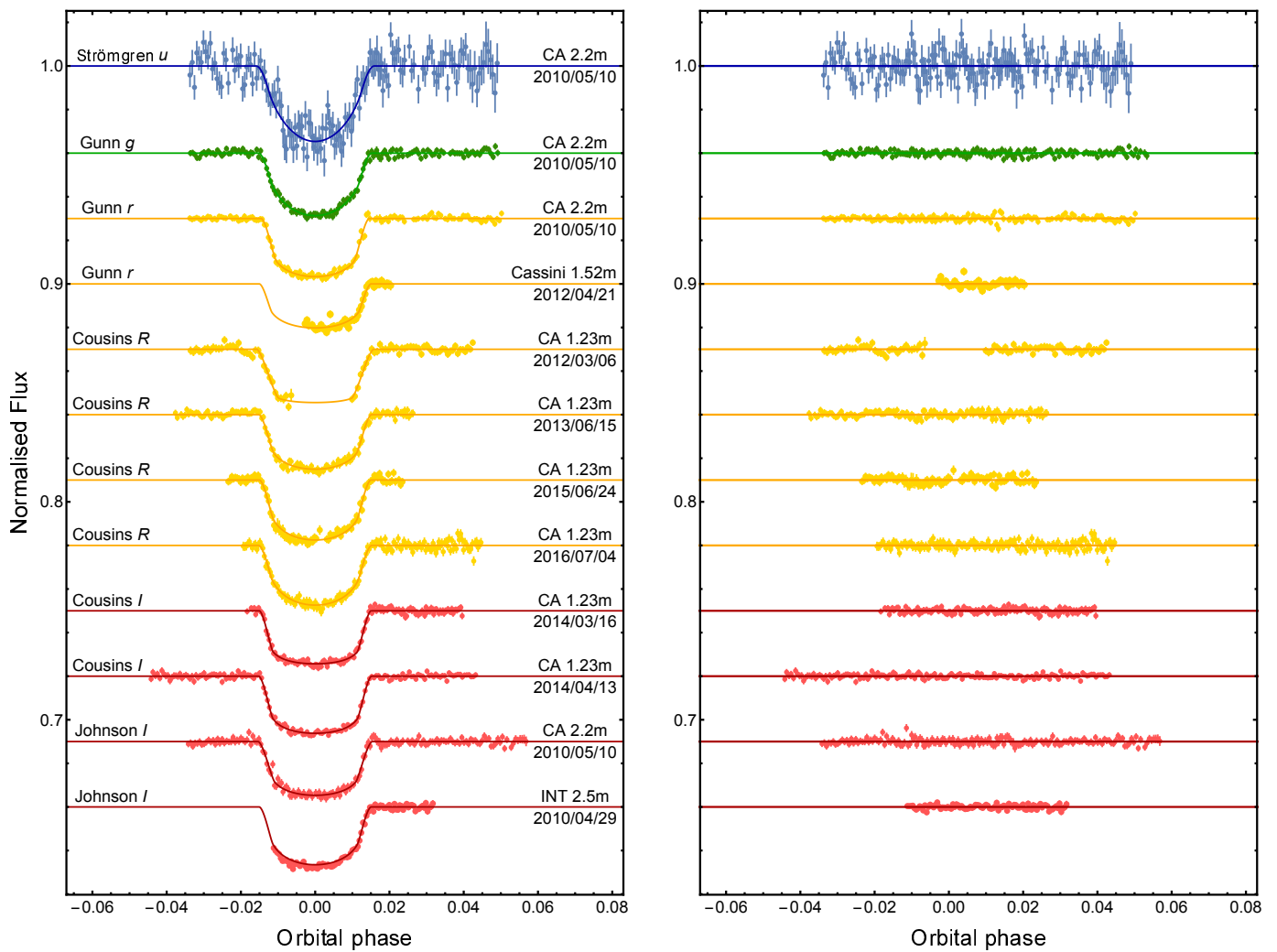}
\caption{Phased light curves of HAT-P-12\,b transits presented in this work. These phased light curves are ordered based on the filter used and compared with the best {\sc jktebop} fits. The dates, telescopes, and filters related to the observation of each transit event are indicated. Residuals from the fits are plotted in the {\it right-hand panel}.
}
\label{fig:hatp12_lc}
\end{figure*}
%
\begin{figure}
\centering
\includegraphics[width=\hsize]{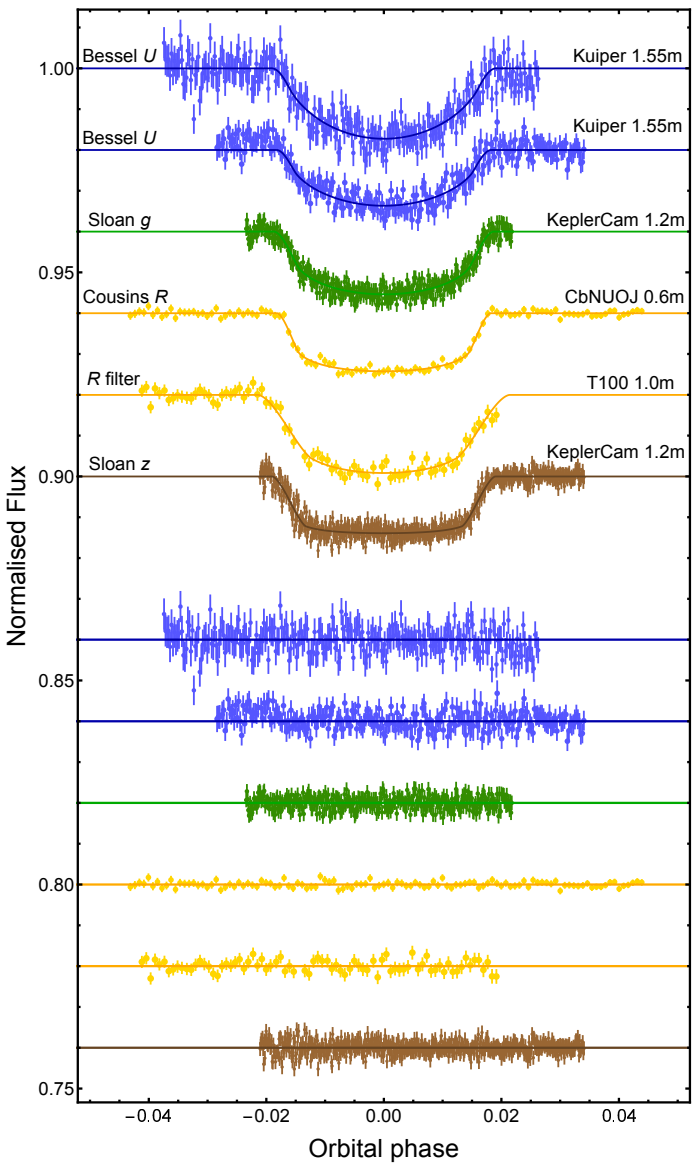}
\caption{Phased light curves of HAT-P-22\,b transits taken from the literature. These phased light curves are compared with the best {\sc jktebop} fits. The telescopes and filters related to the observation of each transit event are indicated. Residuals from the fits are plotted at the base of the figure.}
\label{fig:hatp22_lc}
\end{figure}
%
\begin{figure*}
\centering
\includegraphics[width=18cm]{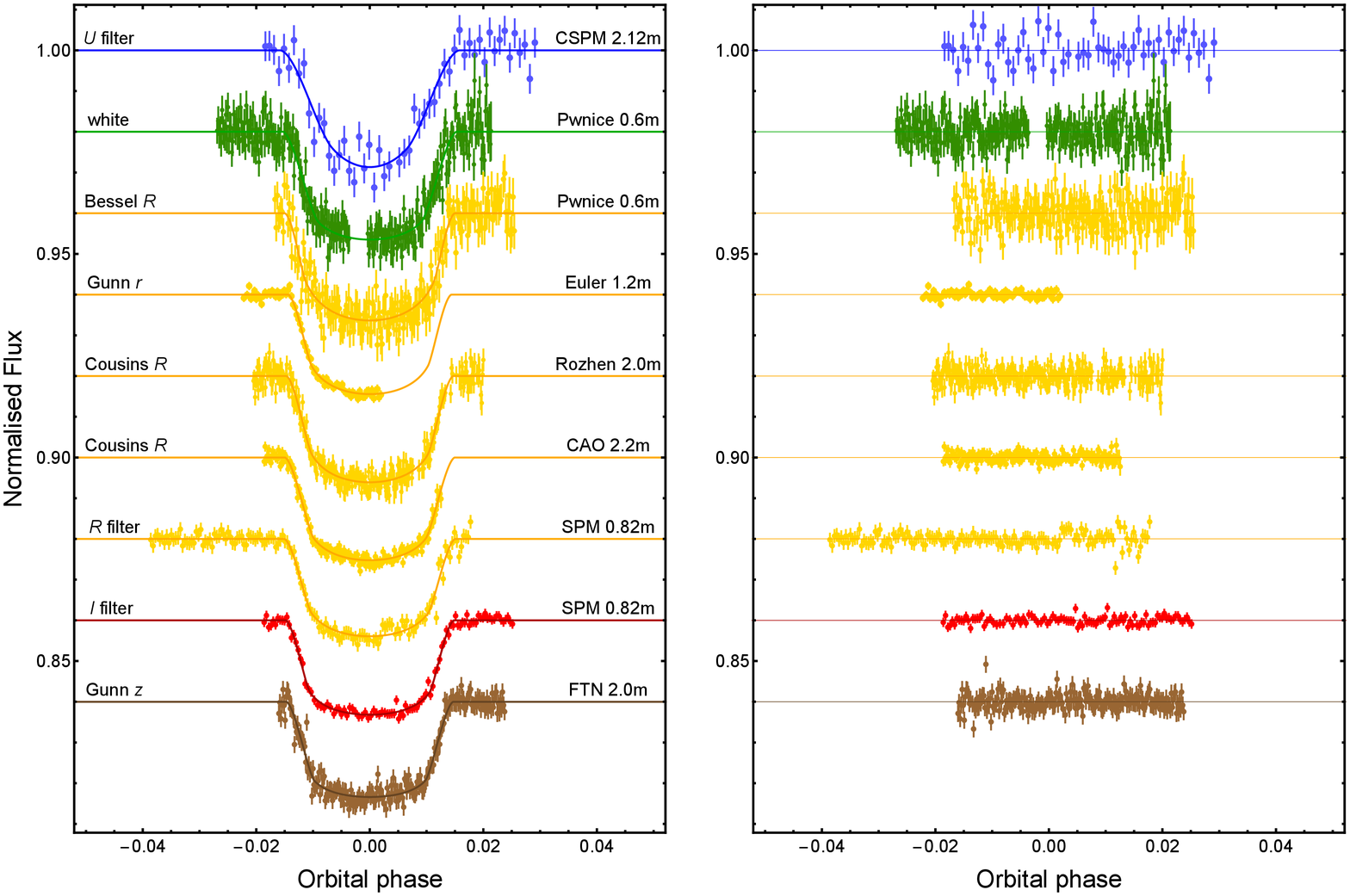}
\caption{Phased light curves of WASP-39\,b transits taken from the literature. These phased light curves are compared with the best {\sc jktebop} fits. The telescopes and filters related to the observation of each transit event are indicated. Residuals from the fits are plotted in the {\it right-hand panel}.}
\label{fig:wasp39_lc}
\end{figure*}
%
\begin{figure}
\centering
\includegraphics[width=\hsize]{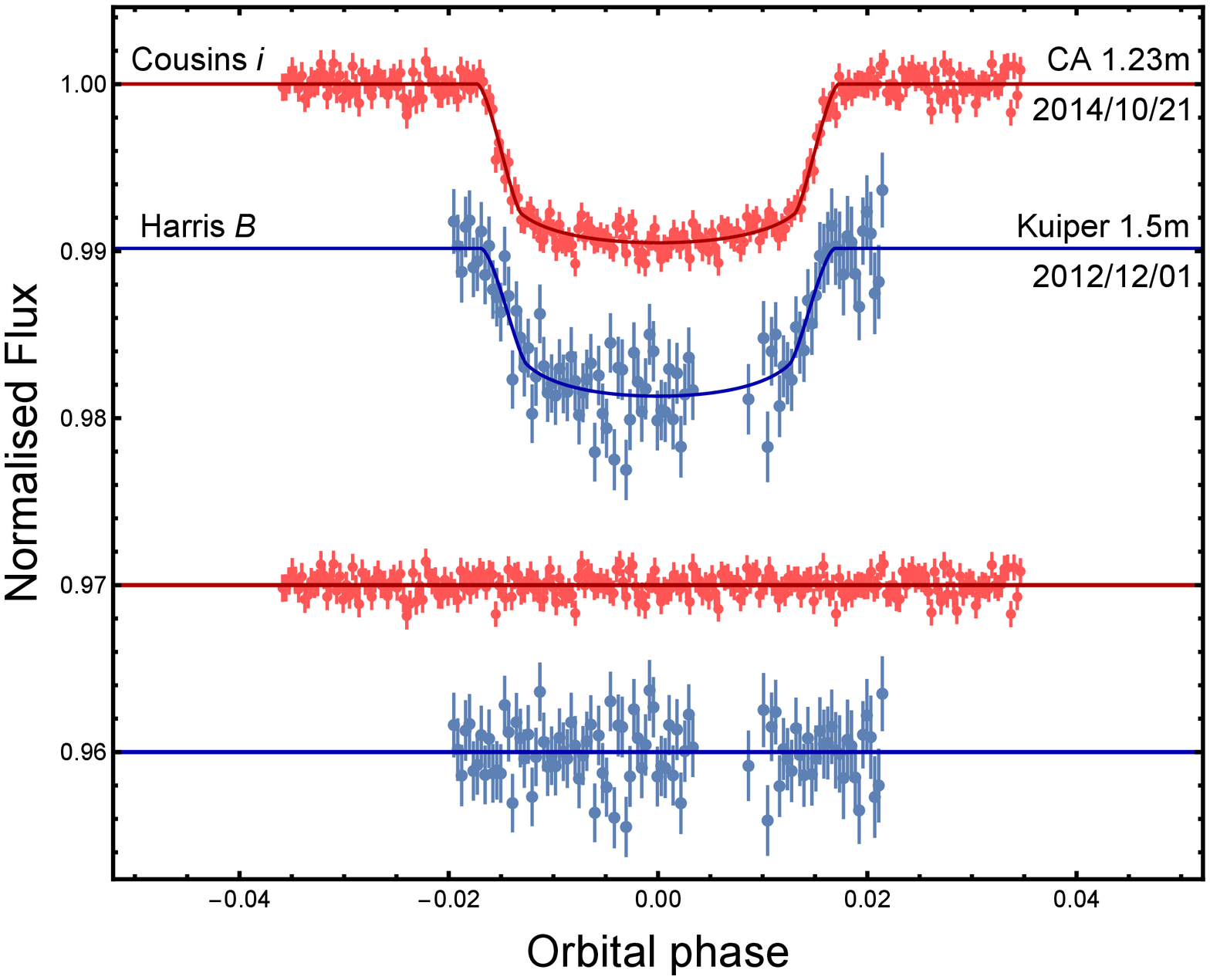}
\caption{{\it Top light curve}: Phased light curve of the WASP-60\,b transit presented in this work. {\it Bottom light curve}: Phased light curve of a WASP-60\,b transit observed by \citet{turner:2017}. The dates, telescopes, and filters related to the observation of these two transit events are indicated. Both light curves are compared with the best {\sc jktebop} fits and the corresponding residuals are plotted at the base of the figure.}\label{fig:wasp60_lc}
\end{figure}

\subsection{Photometric follow-up observations}
\label{sec:photometric_observation}
Three of the planetary systems studied in this work were monitored with an array of medium-class telescopes with the purpose of obtaining high-quality light curves, which are extremely useful for refining the physical parameters and checking stellar activity. These systems are HAT-P-3, HAT-P-12, and WASP-60; the telescopes used were the 2.5 m Isaac Newton Telescope (INT) in La Palma (Spain), the Cassini 1.52\,m Telescope at the Astronomical Observatory of Bologna in Loiano (Italy), and the 2.2\,m and 1.23\,m telescopes at the German-Spanish Astronomical Centre at Calar Alto (Spain). As in previous observations with these telescopes (e.g. \citealp{mancini:2013,mancini:2015}), the defocussing technique was adopted in all the observations to improve the quality of the photometric data significantly \citep{southworth:2009a}. Telescopes were also autoguided and the corresponding CCDs were used unbinned. Sets of flat-fields frames were taken by observing blank fields during the sunset on the same nights as the transits. Together with bias frames, the flat fields were used to calibrate the scientific images during the data reduction phase (see Sect.~\ref{sec:photometricdatareduction}). Details of each transit observation are reported in Table~\ref{tab:logs}.

\subsubsection{Photometric follow-up observations of HAT-P-3}
\label{sec:hatp3_observation}
A complete transit of HAT-P-3\,b was observed on June 2010 through a Gunn-$i$ filter with the BFOSC (Bologna Faint Object Spectrograph \& Camera) imager, which is mounted on the Cassini 1.52\,m Telescope. The BFOSC is equipped with a back-illuminated CCD with $1300\times1340$\,pixels and a pixel size of $20\,{\rm \mu}$m. A focal reducer makes the telescope a f/5, implying a plate scale of 0.58\,arcsec\,pixel$^{-1}$ and a field of view (FOV) of 13\,arcmin $\times$ 12.6\,arcmin.

Three successive complete transits of HAT-P-3\,b were observed between May and June 2015 using the Calar Alto (CA) Zeiss 1.23\,m telescope. This telescope is equipped with the DLR-MKIII camera, which has $4000 \times 4000$ pixels of $15\,\mu$m size. The plate scale is 0.32\,arcsec\,pixel$^{-1}$, which gives a FOV of 21.5\,arcmin $\times$ 21.5\,arcmin. All the three transits were observed using a Cousins-$I$ filter.

\subsubsection{Photometric follow-up observations of HAT-P-12}
\label{sec:hatp12_observation}
A partial transit of HAT-P-12\,b was observed at the end of April 2010 using the wide field camera (WFC) at the prime focus of the INT 2.5\,m telescope. WFC consists of four thinned EEV 2k$\times$4k CCDs, which have a pixel size of $13.5\,\mu$m corresponding to 0.33\,arcsec\,pixel$^{-1}$. A complete transit was observed few nights later using the CA 2.2\,m telescope and the multi-band imager BUSCA. This instrument is equipped with dichroics, which split the incoming starlight towards four Loral CCD4855 cameras ($4000 \times 4000$ pixels of $15\,\mu$m size), allowing simultaneous broadband, four-band transit photometry in the optical window \citep{southworth:2012b,mancini:2014,ciceri:2015}. For this transit, we chose to have Str\"{o}mgren-$u$ filter in the bluest arm, Gunn-$g$ and $r$ filters at intermediate bands, and Johnson $I$ in the reddest arm.

Another partial transit was observed through a Gunn-$r$ filter with the Cassini telescope on April 2012. By using the CA 1.23\,m telescope we observed other six (five complete and one partial) transit events of HAT-P-12\,b. These observations were performed between 2012 and 2016 using Cousins-$R$ (four times) and $I$ (two times) filters.

\subsubsection{Photometric follow-up observations of WASP-60}
\label{sec:wasp60_observation}
A transit of WASP-60\,b was observed on October 2014 through a Cousins-$I$ filter with the CA 1.23\,m telescope. To our knowledge, this is the only complete follow-up transit event that was ever observed for this target.

\subsection{Reduction of the photometric data}
\label{sec:photometricdatareduction}
The photometric data were reduced using a modified version of the {\sc defot} pipeline. This is written in IDL\footnote{The acronym IDL stands for Interactive Data Language and is a trademark of Harris Geospatial Solutions.} and described by \citet{southworth:2014}. Briefly summarising the procedure, we made a median combination of all the calibration images to create master-bias and master-flat frames, and we used these to correct the scientific images. We then identified the target and a suitable set of non-variable stars in each scientific image. We placed three apertures around the stars with radii chosen based on the lowest scatter achieved when compared with a fitted model. Pointing variations of the stars were also corrected with respect to a reference image by re-centring the apertures. The non-variable stars were used as reference stars to extract the photometry of the target using the {\sc aper} routine\footnote{APER is part of the ASTROLIB subroutine library distributed by NASA.}. Light curves were created for each transit data set with a first or a second-order polynomial fitted to the out-of-transit data. We simultaneously fit the comparison-star weights and polynomial coefficients to minimise the scatter outside the transit. The final light curves are plotted in Figs. \ref{fig:hatp3_lc}, \ref{fig:hatp12_lc}, and \ref{fig:wasp60_lc}.

\section{Light-curve analysis}
\label{sec:lc_analysis}

\begin{table*}
\caption{Final photometric parameters for the five exoplanetary systems analysed in this work. The parameters $r_{\star}$ and $r_{\rm p}$ are the fractional stellar and planetary radius, respectively. The quantities in brackets denote the uncertainty in the final digit of the preceding number.}
\label{tab:photparameters}
\centering
\begin{tabular}{c c c c c c }
\hline\hline \\ [-6pt]
System ~~& Orbital period & Time of mid-transit & Orbital inclination, & $r_{\star}+r_{\rm p}$ & $r_{\rm p}/r_{\star}$\\
& (days) & (BJD$-2400000$) & $i$ (degrees) &  & \\
\hline \\ [-6pt]
HAT-P-3 \tablefootmark{a}~~ & 2.89973838\,(27) & 57150.39472\,(58) & $86.31 \pm 0.19$ & $0.11317 \pm 0.00180$ & $0.11056 \pm 0.00068$ \\
HAT-P-12\tablefootmark{b}   & 3.21305992\,(35) & 55328.49068\,(19) & $89.10 \pm 0.24$ & $0.09549 \pm 0.00095$ & $0.13898 \pm 0.00069$ \\
HAT-P-22\tablefootmark{c}   & 3.21223328\,(58) & 54930.22016\,(16) & $86.46 \pm 0.41$ & $0.13107 \pm 0.00354$ & $0.10911 \pm 0.00065$ \\
WASP-39\tablefootmark{d}    & 4.0552941\,~(34) & 55342.96913\,(63) & $87.32 \pm 0.17$ & $0.10303 \pm 0.00156$ & $0.14052 \pm 0.00077$ \\
WASP-60\tablefootmark{e}    & 4.3050040\,~(59) & 56952.43264\,(17) & $86.05 \pm 0.57$ & $0.12795 \pm 0.00609$ & $0.08986 \pm 0.00009$ \\
\hline
\end{tabular}
\tablefoot{
\tablefoottext{a}{The photometric parameters of HAT-P-3 were estimated from the light curves presented in this work (Fig.~\ref{fig:hatp3_lc}), incorporating results from \citet{southworth:2012} (see text).%
\tablefoottext{b}{The photometric parameters of HAT-P-12 were estimated from the light curves presented in this work, see Fig.~\ref{fig:hatp12_lc}.}
\tablefoottext{c}{The photometric parameters of HAT-P-22 were estimated from the light curves taken from various works \citep{bakos:2011,basturk:2015,hinse:2015,turner:2016}, see Fig.~\ref{fig:hatp22_lc}.}
\tablefoottext{d}{The photometric parameters of WASP-39 were estimated from the light curves taken from different works \citep{faedi:2011,ricci:2015,maciejewski:2016}, see Fig.~\ref{fig:wasp39_lc}.}
\tablefoottext{e}{The photometric parameters of WASP-60 were estimated from the light curve presented in this work and the one from \citet{turner:2017}, see Fig.~\ref{fig:wasp60_lc}.}
}}
\end{table*}

The light curves of the transit events of HAT-P-3, HAT-P-12, and WASP-60, which were presented in the previous section, were individually studied with the {\sc jktebop} code \citep{southworth:2013} to find the best-fitting model for each of these events. 
For HAT-P-3, we also considered the {\sc jktebop} best-fitting results obtained by \citet{southworth:2012}, who analysed five published light curves of HAT-P-3\,b transits;
moreover, for WASP-60, we considered the partial light curve obtained by \citet{turner:2017} and modelled this event with {\sc jktebop} as well (see Fig~\ref{fig:wasp60_lc}).

For the other two systems, HAT-P-22 and WASP-39, we collected all the light curves available from the literature and modelled each one of these with {\sc jktebop} as well. For HAT-P-22 we used the light curves from \citet{bakos:2011}, \citet{basturk:2015}, \citet{hinse:2015}, and \citet{turner:2016}; while for WASP-39 we used those from \citet{faedi:2011}, \citet{ricci:2015}, and \citet{maciejewski:2016}. These light curves are shown in Figs.~\ref{fig:hatp22_lc} and \ref{fig:wasp39_lc}, in which the telescopes and the filters used are also specified.

The {\sc jktebop} code represents the star and planet as spheres and uses the Levenberg-Marquardt optimisation algorithm to fit the parameters of the light curves. The main parameters to be fitted are the orbital period and inclination ($P$ and $i$),  time of transit midpoint ($T_0$), and sum and ratio of the fractional radii, which are defined as $r_{\star}=R_{\star}/a$ and $r_{\rm p}=R_{\rm p}/a$, where $R_{\star}$ and $R_{\rm p}$ are the true radii of the star and planet, and $a$ is the semi-major axis of the planetary orbit. We used a quadratic law to describe the effect of the limb darkening (LD) of the star and fitted the LD coefficients with {\sc jktebop}, taking into account the differences between the properties of the various stars and filters used. The orbital eccentricity was fixed to zero for all the systems, based on the results of \citet{bonomo:2017}. Since time-series photometry is generally affected by correlated (red) noise \citep{carter:2009}, which is not taken into account by the {\sc aper} routine, we inflated the error bars of the photometric measurements to give a reduced $\chi^2$ of $\chi^2_{\nu}=1$ during the best-fitting process of each light curve. The light curves and corresponding {\sc jktebop} best-fitting lines are reported in Fig.~\ref{fig:hatp3_lc}, \ref{fig:hatp12_lc}, \ref{fig:hatp22_lc}, \ref{fig:wasp39_lc}, and \ref{fig:wasp60_lc} for HAT-P-3, HAT-P-12, HAT-P-22, WASP-39, and WASP-60, respectively.

The uncertainties of the fitted parameters were also estimated with {\sc jktebop}, by running both a Monte Carlo and a residual-permutation algorithm. For each light curve, we ran at least 10\,000 simulations for the Monte Carlo algorithm, and the maximum number of simulations (i.e.\ one less than the number of datapoints) for the residual-permutation algorithm, and adopted the largest of the two 1$\sigma$ values as the final uncertainty for each parameter. The final values of each parameter were finally estimated by means of a weighted average of the values extracted from the fit of all the individual light curves, using the relative uncertainties as a weight, and these values are reported in Table~\ref{tab:photparameters}. The orbital ephemerides are also shown, as they were recalculated performing a weighted linear least-squares fit to all the mid-transit times versus their cycle number. For this task, we considered all the light curves that were discussed above and the times of mid-transit from the discovery papers.

\section{Frequency analysis of the time-series light curves}
\label{sec:frequency_analysis}
Time-series photometric data are available in the WASP\footnote{https://exoplanetarchive.ipac.caltech.edu} and HAT\footnote{https://hatnet.org} databases for the five stars included in this study. These data are very dense (thousands of measurements) and span a long time baseline (hundreds of days). They can be useful for detecting any periodic or quasi-periodic signal, which could indicate stellar activity and hence allow us to suggest a rotational period for some of the five stars. We used the iterative sine-wave least-squares method \citep{vani} to perform the frequency analysis. We also obtained amplitude spectra and we determined the mean level noise in the 0.01-0.90~d$^{-1}$ frequency interval. We then computed the S/N of each peak to infer the significance by assuming a threshold of 4.0 \citep{snr4}

\subsection{Stars without a clear signal: WASP-60, HAT-P-12, WASP-39}
The frequency analysis of the WASP-60 time series did not detect any  peak. The mean level of the noise is 0.29~mmag.
The analysis of the data collected on HAT-P-12 does not suggest a clear value for the rotational frequency, although some structures of peaks can be occasionally seen in the spectra obtained from the different photometry provided. The noise level of the HAT-P-12  measurements is 0.89~mmag and this hampers  the  clear identification of the rotational signals, if any.

The frequency spectrum of the time series on WASP-39 shows a peak that is close to the frequency corresponding to the synodic month. On the basis of the photometric data only, we conservatively interpret such a peak as spurious, since WASP-39 is an equatorial star and hence moonlight can alter the measurements in an almost regular way.

\subsection{Stars with a  signal: HAT-P-3, HAT-P-22}
The original time series of HAT-P-3 is very noisy and characterised by a dense group of outliers. By removing these outliers, we obtained a less scattered data set showing  a noise level of 1.1~mmag. The frequency analysis reveals a peak at 0.054~d$^{-1}$, corresponding to 18.5014~d. The amplitude is 3.8~mmag, putting the ${\rm S/N}=3.4$ detection below the significance level.

\begin{figure}
\centering
\includegraphics[width=9cm]{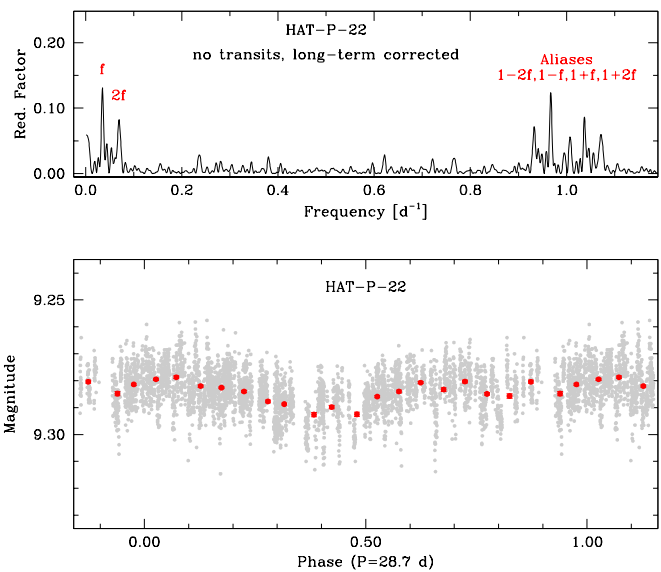}
\caption{Detection of the rotational period of HAT-P-22. {\it Top:} Power spectrum of the photometric measurements. The peaks corresponding to $f$=0.0345~d$^{-1}$, $2f$, and their aliases are indicated. {\it Bottom:} The photometric measurements (in grey) folded with $P$=28.7~d. The error bars of the mean values of the binned data (red circles) have the same size of the points.}
\label{hat22}
\end{figure}

The case of HAT-P-22 is by far the most interesting. The preliminary periodogram showed a peak at a very low frequency, which is produced by a long-term trend in the time series. We corrected it by subtracting a linear best fit and recomputed the periodogram. In such a way, we could clearly detect two peaks at 0.0345 and 0.0690~d$^{-1}$ (Fig.~\ref{hat22}, top panel). These two peaks are not related to any peak in the spectral window of the data. The second frequency is twice the first. The amplitude of this signal is 5.6~mmag, that of the noise is 0.8~mmag only (${\rm S/N}=7.1$, largely significant). We note that $f=0.0345$~d$^{-1}$ is still a frequency close to that corresponding to the synodic month, but unlike WASP-39, HAT-P-22 is a high-declination star and the signal shows a larger amplitude and a highly non-sinusoidal shape. Therefore, we are keen to interpret such detections as a clear hint of a flux modulation over a rotational period close to 29~d (corresponding to $f=0.0345$~d$^{-1}$), shaped like a double wave by the harmonic $2f$ (Fig.~\ref{hat22}, bottom panel). The low-frequency peak can be ascribed by long-term variations of the surface features, due to a much longer stellar activity cycle.
We refined the value of the rotational period using the MTRAP code \citep{carpino:1987}, allowing the simultaneous fit with $f, 2f,$ and the low-frequency component, thus obtaining $P_{\rm rot}=28.7\pm0.4$~d. 

\section{HARPS-N spectra analysis}
\label{sec:spectra_analysis}

\subsection{Stellar atmospheric parameters}
\label{sec:stellar_parameters}

We used the HARPS-N spectra to perform a detailed spectroscopic characterisation of the host-star atmospheric parameters, i.e. effective temperature $T_{\rm eff}$, surface gravity $\log{g}$, iron abundance [Fe/H], and projected rotational velocity $v\sin{i_{\star}}$, where $i_{\star}$ is the inclination of the stellar rotation axis with respect to the line of sight. To this purpose, for each target we obtained a single, high S/N, 1-D spectrum by averaging all the spectra available, after correcting each one spectrum for the corresponding RV shift; these spectra, acquired during bright-time, were corrected for moonlight contamination by subtracting the sky-background estimated from fibre B, as described in Sect.~\ref{sec:HARPS-N_observation}.

Preliminary estimates of the effective temperature for the five targets were obtained by applying the method of equivalent width (EW) ratios of photospheric absorption lines, making use of the ARES\footnote{http://www.astro.up.pt/$\sim$sousasag/ares/} automatic code \citep{sousa:2007} and the calibration for FGK dwarf stars by \citet{sousa:2010}. The atmospheric stellar parameters, $T_{\rm eff}$, $\log{g}$, and [Fe/H] were then derived via the program MOOG (\citealp{sneden:2007}; version 2013) and EW measurements of iron lines, as described in detail by \citet[and references therein]{biazzo:2012}. The projected rotational velocity was estimated with the same code and applying the spectral synthesis method \citep{dorazi:2011}. The results are reported in Table\,\ref{tab:star-par} and are in good agreement with previous estimates, with the exception of WASP-60, for which we measured a slightly hotter temperature and a higher iron abundance (see Sect.~\ref{sec:physical_parameters}).

\begin{table}
\small{
\caption{Stellar atmospheric parameters determined from HARPS-N spectra.}
\label{tab:star-par}
\centering
\begin{tabular}{c c c c c}     
\hline\hline \\ [-6pt]
Object & $T_{\rm eff}$ & $\log{g}$ & [Fe/H] & $v \sin{i_{\star}}$\\
& (K) & (dex) & & (km\,s$^{-1}$) \\
\hline \\ [-6pt]
HAT-P-3~~ & $5190 \pm 80$ & $4.58 \pm 0.10$ & $+0.24 \pm 0.08$    & $1.4 \pm 0.5$ \\ %
HAT-P-12  & $4665 \pm 50$ & $4.55 \pm 0.21$ & $-0.20 \pm 0.09$~~~ & $0.5 \pm 0.5$ \\ %
HAT-P-22  & $5314 \pm 50$ & $4.39 \pm 0.16$ & $+0.30 \pm 0.09$    & $1.3 \pm 0.7$ \\ %
WASP-39   & $5485 \pm 50$ & $4.41 \pm 0.15$ & $+0.01 \pm 0.09$    & $1.0 \pm 0.5$ \\ %
WASP-60   & $6105 \pm 50$ & $4.31 \pm 0.11$ & $+0.26 \pm 0.07$    & $3.8 \pm 0.6$ \\ %
\hline
\end{tabular}
}
\end{table}

\subsection{Stellar activity indexes}
\label{sec:activity_index}
The average spectra were also used to analyse the Ca\,II H\&K lines and measure both the chromospheric Mount Wilson S-index and $\log{R^{\prime}_{\rm HK}}$ index for each of the five stars; see Table~\ref{tab:activity_index}. In particular the $\log{R^{\prime}_{\rm HK}}$ indexes indicate low activity in all the cases, confirming the general non-detections of the rotational periods in the photometric data (see Sect.~\ref{sec:frequency_analysis}). Adopting the calibration scales by \citet{noyes:1984} and \citet{mamajek:2008}, the level of the stellar activity provides an indication of how fast the stars rotate and how old they are. The projected rotation velocity and age estimated in this way are also reported in Table~\ref{tab:activity_index}.
We note that the predicted $P_{\rm rot}$ for HAT-P-22 (48-52~d) seems too long with respect to that determined from photometric data (28.7~d). Also the values of $P_{\rm rot}$ for HAT-P-12 and WASP-39 are not those expected for mid/late K-dwarf stars \citep{McQuillan:2014}. On the other hand, the predicted $P_{\rm rot}$ for HAT-P-3 (20~d) is in good agreement with the value suspected from photometry (18.5~d).

\begin{table*}
\small{
\caption{Stellar activity indexes and related parameters.}
\label{tab:activity_index}
\centering
\begin{tabular}{c c c c c c c}     
\hline\hline \\ [-6pt]
Object & S-index & $B-V$ & $\log{R^{\prime}_{\rm HK}}$ & ~~$P_{\rm rot}$\tablefootmark{(a)} & ~~$P_{\rm rot}$\tablefootmark{(b)} & ~~~Age\tablefootmark{(c)} \\
 & & & & (days) & (days) & Gyr \\
\hline \\ [-6pt]
HAT-P-3~~ & $0.217 \pm 0.016 $ & 0.67 & $-4.75 \pm 0.06$ & $20.2 \pm 2.0$ & $19.6 \pm 2.3$ & $2.6 \pm 0.6$  \\ %
\hline \\ [-6pt]
~~~~HAT-P-12\tablefootmark{(d)}  &
$\begin{tabular}{c}
$0.364 \pm 0.054$ \\
$0.375 \pm 0.075$ \\
\end{tabular}$ & 1.09 &
$\begin{tabular}{c}
$-4.88 \pm 0.07$ \\
$-4.87 \pm 0.09$ \\
\end{tabular}$ &
$\begin{tabular}{c}
$43.9 \pm 3.7$ \\
$43.3 \pm 5.0$ \\
\end{tabular}$ &
$\begin{tabular}{c}
$44.5 \pm 4.9$ \\
$43.9 \pm 6.5$ \\
\end{tabular}$ &
$\begin{tabular}{c}
$5.5 \pm 1.1$ \\
$5.3 \pm 1.4$ \\
\end{tabular}$ \\
\hline \\ [-6pt]
HAT-P-22  & $0.154 \pm 0.004 $ & 0.86 & $-5.09 \pm 0.02$ & $48.1 \pm 0.8$ & $52.6 \pm 1.1$ & $9.6 \pm 0.4$ \\ %
\hline \\ [-6pt]
WASP-39   & $0.183 \pm 0.017 $ & 0.84 & $-4.97 \pm 0.06$ & $42.1 \pm 2.6$ & $44.0 \pm 3.9$ & $7.2 \pm 1.1$ \\ %
\hline \\ [-6pt]
WASP-60   & $0.146 \pm 0.009 $ & 0.68 & $-5.10 \pm 0.07$ & $31.8 \pm 1.9$ & $34.8 \pm 2.7$ & $6.84 \pm 0.93$ \\ %
\hline
\end{tabular}
\tablefoot{
\tablefoottext{a}{This value was obtained adopting \citet{noyes:1984} calibration scale.}%
\tablefoottext{b}{This value was obtained adopting \citet{mamajek:2008} calibration scale.}
\tablefoottext{c}{This value was obtained adopting \citet{mamajek:2008} calibration scale.}
\tablefoottext{d}{The top values refer to the transit observed on 2015.03.14, while the bottom values to the transit observed on 2015.04.24.}
}}
\end{table*}

\subsection{Determination of the spin-orbit alignment}
\label{sec:alignment}

\begin{table*}
\caption{Parameters from the best-fitting models of the RM effect for the five planetary systems. The $\Delta$BIC values, i.e. the difference between the BIC values calculated from the best-fitting models with and without the RM effect, are also shown.}
\label{tab:RM}
\centering
\begin{tabular}{c c c c c}
\hline\hline \\ [-6pt]
Object & $\lambda$ & $v \sin{i_{\star}}$ & $\gamma$ & $\Delta$BIC    \\
       & (degree)  & (km\,s$^{-1}$)      & (km\,s$^{-1}$)  \\
\hline \\ [-6pt]
HAT-P-3 \tablefootmark~~ & $\,21.2 \pm 8.7$   & $1.20 \pm 0.36$ & $-23.3849 \pm 0.0007$ & 33.5\\ %
HAT-P-12\tablefootmark   & $-54^{+41}_{-13}$ & $0.99^{+0.42}_{-0.46}$ & $-40.4589 \pm 0.0023$ & 10.3\\ %
HAT-P-22\tablefootmark   & $-2.1 \pm  3.0$   & $1.65 \pm 0.26$ & $ ~\,\,12.6370 \pm 0.0004$ & 71.4\\ %
WASP-39\tablefootmark    & $\,\,\,\,\,\,\,0\pm 11  $   & $1.40 \pm 0.25$ & $-58.4421 \pm 0.0020$ & 19.7\\ %
WASP-60\tablefootmark    & $-129 \pm 17~~$   & $2.97 \pm 0.47$ & $-26.5323 \pm 0.0021$ & 19.3\\ %
\hline
\end{tabular}
\end{table*}

The analysis of the HARPS-N RV data, for measuring the orbital obliquity of the five planetary systems, was performed using a code developed by our team, which was already used for this purpose in the previous works of the series. The most recent work, \citet{esposito:2017}, provides  a detailed description of the RM-effect modelling and the fitting algorithm.

\begin{figure*}
\centering
\includegraphics[width=18cm]{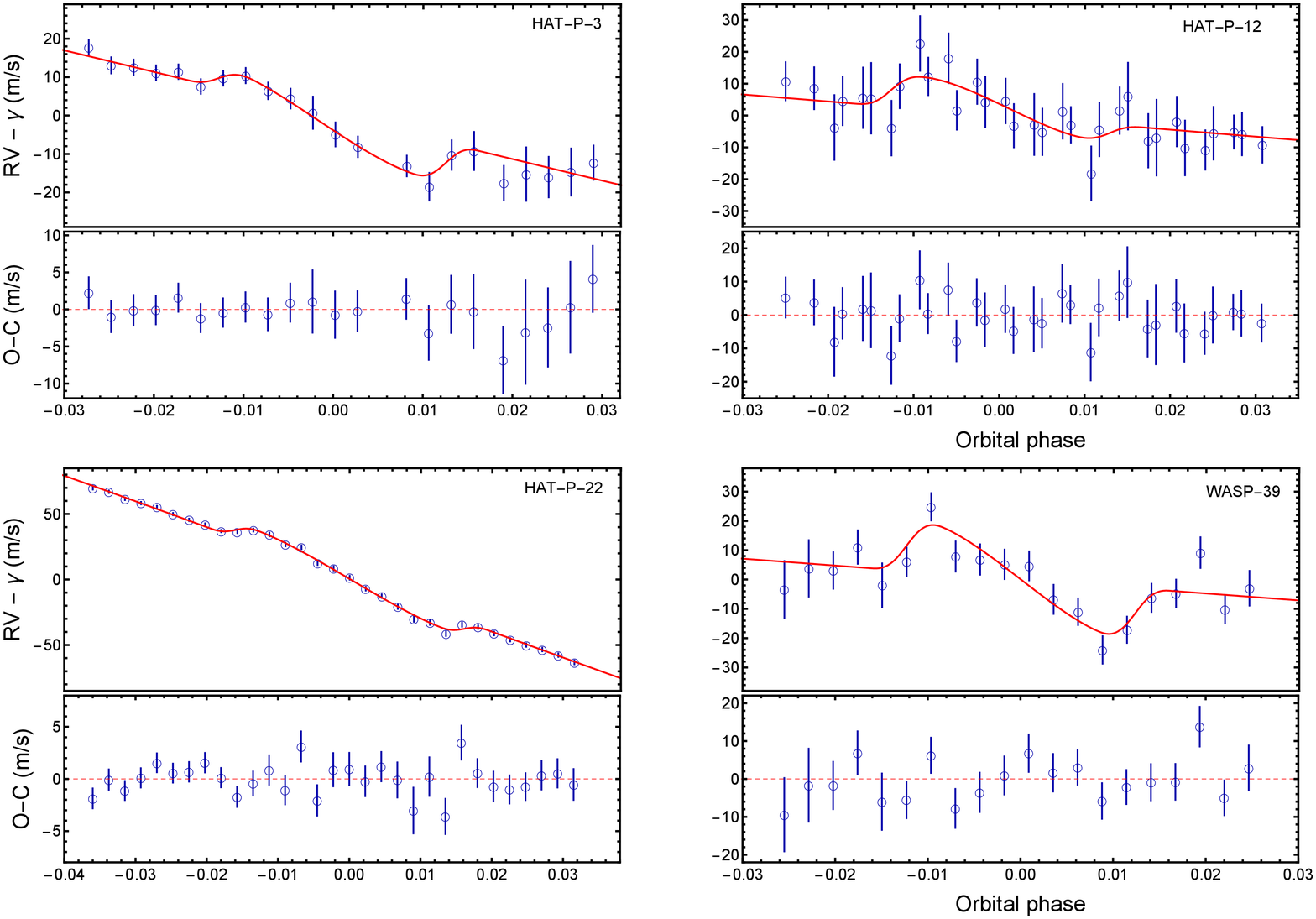}
\caption{Phase-folded RV data of HAT-P-3, HAT-P-12, HAT-P-22, and WASP-39 taken with HARPS-N during planetary-transit events. Superimposed are the best-fitting RV-curve models. The corresponding residuals are plotted in the bottom panels.}
\label{fig:RM_plot}
\end{figure*}

\begin{figure}
\centering
\includegraphics[width=\hsize]{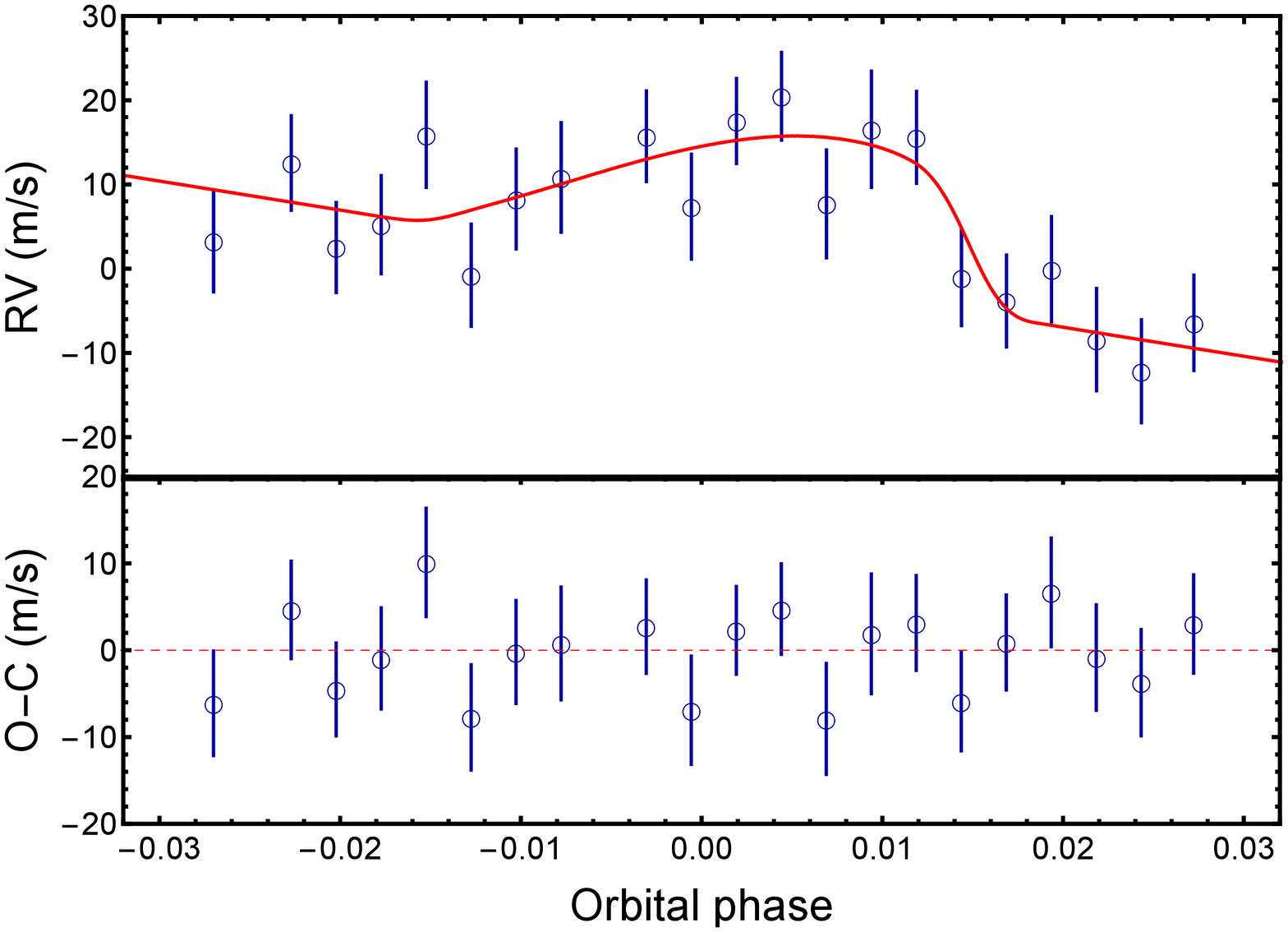}
\caption{Phase-folded RV data of WASP-60 taken with HARPS-N during a planetary-transit event. Superimposed is the best-fitting RV-curve model. The corresponding residuals are plotted in the bottom panel.}
\label{fig:wasp60_RM_plot}
\end{figure}

We used the transit-bracketing RV time series to derive the best-fitting values for three parameters: the sky-projected orbital obliquity angle $\lambda$, the stellar projected rotational velocity $v\sin{i_\star}$, and the systemic RV $\gamma$. All the other relevant parameters were kept fixed to the values obtained by the photometric and spectroscopic data analysis, and their uncertainties were propagated in the determination of the error bars for $\lambda$, $v\sin{i_\star}$, and $\gamma$. The final results are reported in Table~\ref{tab:RM}, while the best-fitting RV models are shown in Figs.~\ref{fig:RM_plot} and \ref{fig:wasp60_RM_plot}, superimposed on the data sets. We found that HAT-P-22 and WASP-39 are planetary systems with well-aligned orbits ($\lambda=-2.1^{\circ}\pm 3.0^{\circ}$ and $\lambda=0^{\circ}\pm 11^{\circ}$, respectively); the orbit of HAT-P-3\,b is slightly misaligned ($\lambda=21.2^{\circ}\pm 8.7^{\circ}$); concerning HAT-P-12\,b, our results also indicate a very misaligned orbit, however this is not well constrained ($\lambda=-54^{\circ}\,^{+41^{\circ}}_{-13^{\circ}}$); finally, WASP-60\,b clearly shows a retrograde orbit ($\lambda=-129^{\circ} \pm 17^{\circ}$), Fig.~\ref{fig:wasp60_RM_plot}.

Considering the small amplitudes of the RM effects and the accuracy of our RV measurements, we used the Bayesian information criterion (BIC) to perform a statistical comparison of our best-fitting results with those without RM effect. The BIC value was calculated for each data set using the corresponding number of data points, for both cases, i.e. with and without RM (the number of parameters estimated from the model was 3 for the fit with the RM effect and 1 for the fit with no RM effect). The differences in BIC are reported in Table~\ref{tab:RM} and they strongly support, in all cases, the statistical significance of the detection of the RM effect.

Knowing the rotational period of HAT-P-22 ($P_{\rm rot}=28.7 \pm 0.4$\,d; see Sect.~\ref{sec:frequency_analysis}), we used the following formula:
\begin{equation}
P_{\mathrm{rot}}\approx \frac{2 \pi
R_{\star}}{v\sin{i_{\star}}}\sin{i_{\star}}, %
\label{Eq:1}
\end{equation}
to estimate the angle $i_{\star}$, which resulted to be $62^{\circ}\pm19.0^{\circ}$. Then, we can simply estimate the true misalignment angle by using Eq. (7) in \citet{winn:2007}, i.e.
\begin{equation}
\cos{\psi}=\cos{i_{\star}}\cos{i}+\sin{i_{\star}}\sin{i}\cos{\lambda}, %
\label{Eq:2}
\end{equation}
thus obtaining $\psi=25^{\circ} \pm 18^{\circ}$.

%
%

\section{Physical parameters}
\label{sec:physical_parameters}
Based on the data described in the previous sections, we reviewed the physical properties of the planetary systems HAT-P-3, HAT-P-12, HAT-P-22, WASP-39, and WASP-60. For this purpose, we followed the {\it homogeneous studies} approach (see \citealp{southworth:2012} and references therein) and combined the measured parameters from the light curves and spectroscopic observations with constraints on the properties
of the host stars coming from theoretical stellar evolutionary models.

The spectroscopic properties of the host stars that we used are the effective temperature $T_{\rm eff}$, the logarithmic surface gravity $\log{g}$, iron abundance, and projected rotational velocity $v \sin{i_{\star}}$. These values were obtained from the analysis of the HARPS-N spectra, Sect.~\ref{sec:spectra_analysis}.

Since most of the HARPS-N data were collected during transit events, we do not have enough out-of-transit RV points for a precise estimation of the velocity amplitude, $K_{\rm A}$, of the RV curves, hence we adopted the values from the literature (see Table~\ref{tab:hatp12_final_parameters}, \ref{tab:hatp22_final_parameters},\ref{tab:wasp39_final_parameters}, \ref{tab:wasp60_final_parameters}). The only exception was the case of HAT-P-3, for which we measured $K_{\rm A}$ using out-of-transit RV HIRES + HARPS-N data.

Having established a good set of input parameters, we used the {\sc jktabsdim} code \citep{southworth:2009} to redetermine the main physical properties of the five planetary systems. The {\sc jktabsdim} code maximises the agreement between the measured $R_{\star}/a$ and $T_{\rm eff}$ with those predicted by a set of five theoretical models by iteratively modifying the velocity amplitude of the planet. The code also considers a wide range of possible ages for each of the host stars and, at the end, returns five different estimates for each of the output parameters. We took the unweighted means as the final values of the parameters. Statistical uncertainties were propagated from the error bars in the values of all input parameters, whereas systematic uncertainties were calculated based on the maximum deviation between the values of the final parameters and individual values coming from the five theoretical models. Our final values are reported in Tables~\ref{tab:hatp3_final_parameters}, \ref{tab:hatp12_final_parameters}, \ref{tab:hatp22_final_parameters}, \ref{tab:wasp39_final_parameters}, and \ref{tab:wasp60_final_parameters}, and compared with values taken from the literature.

For HAT-P-3, HAT-P-12, HAT-P-22, and WASP-39, we found that our measured radii and masses for the stars and planets were all within the error bars of literature determinations. Our results are therefore in good agreement with previous works.

\subsection{Lower density for WASP-60\,b}
\label{sec:wasp60_results}

For WASP-60 we obtained significantly different physical parameters with smaller error bars than in the literature (see Table~\ref{tab:wasp60_final_parameters}). Firstly, we measured a smaller stellar density than in \citet{hebrard:2013}. The latter was based on sparse follow-up photometry. Secondly, based on HARPS-N data, we estimated a higher stellar temperature, i.e. $6105 \pm 50$\,K versus $5900 \pm 100$\,K, which changes the spectral type from G1\,V to F9\,V. Therefore, we found that the WASP-60 is a younger and more massive star than was thought. Moreover, we found that the planet WASP-60\,b is larger ($R_{\rm p}=1.225\pm 0.069\,R_{\rm Jup}$ versus $R_{\rm p}=0.86 \pm 0.12\,R_{\rm Jup}$), much less dense ($\rho_{\rm p}=0.285 \pm 0.052\,\rho_{\rm Jup}$ versus $\rho_{\rm p}=0.8 \pm 0.3\,\rho_{\rm Jup}$), has a smaller surface gravity ($g_{\rm p}=9.2 \pm 1.2$\,m\,s$^{-2}$ versus $g_{\rm p}=15.5^{+4.9}_{-3.7}$\,m\,s$^{-2}$) and is hotter ($T_{\rm eq}=1479 \pm 35$\,K versus $T_{\rm eq}=1320 \pm 75$) than that was measured by \citet{hebrard:2013}.

We stress that, in obtaining these results, we used spectroscopic data of higher quality with respect to those used by \citet{hebrard:2013}. Furthermore, for the analysis of the photometric parameters, we used two light curves: the one presented in Sect.~\ref{sec:wasp60_observation} and the one from \citet{turner:2017}. Comparing the quality of these two light curves (see Fig.~\ref{fig:wasp60_lc}), we note that the light curve obtained with the CA\,1.23\,m telescope has a longer coverage of the baseline (before the ingress and after the egress) and its points are much less scattered and have smaller uncertainties. Therefore, the analysis of the photometric parameters is dominated by the CA light curve, which allowed us to obtain more precise measurements of the contact points and transit depth.

\begin{figure}
\centering
\includegraphics[width=\hsize]{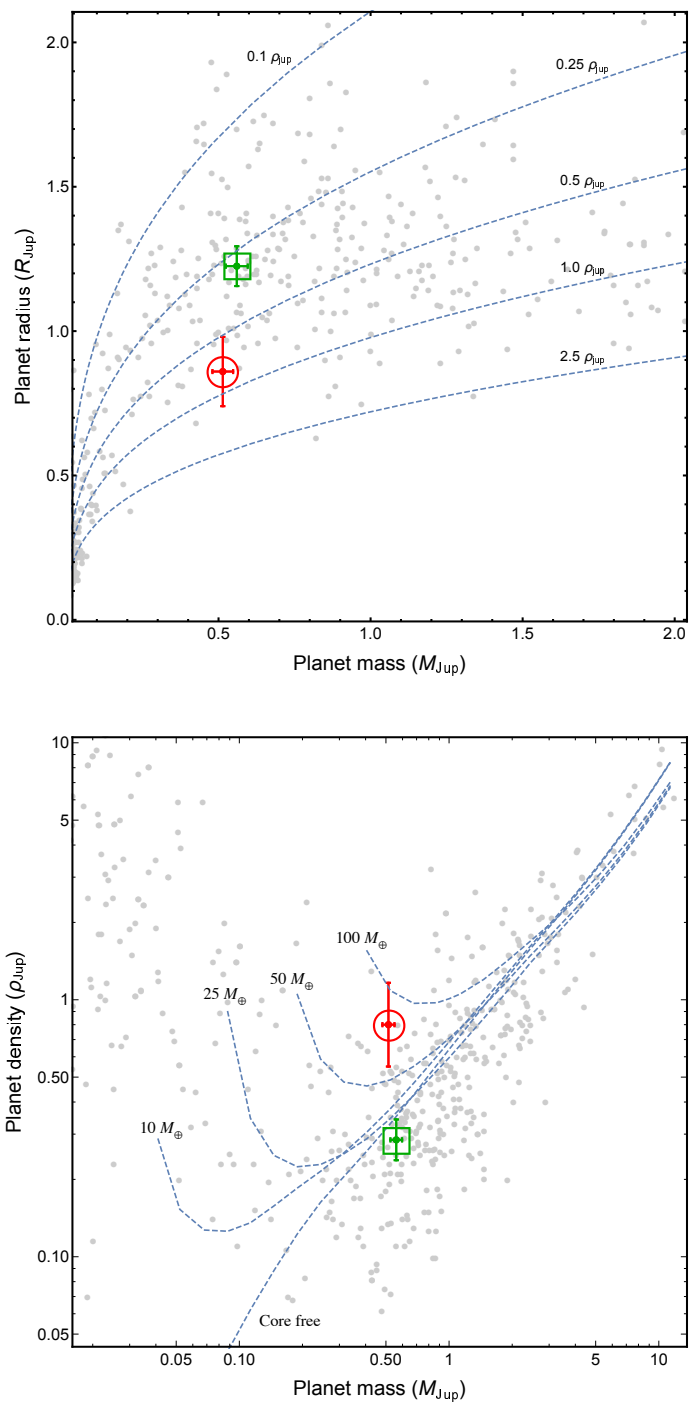}
\caption{Mass-radius and mass-density diagrams for known transiting exoplanets. The positions of WASP-60\,b are shown in both the panels with a green point in a box (this work) and a red point in a circle \citep{hebrard:2013}. The error bars are also illustrated. The grey points denote values taken from TEPCat and their error bars are suppressed for clarity. {\it Top panel}: A zoom of the mass-radius diagram of known transiting exoplanets. Dashed lines show where density is 2.5, 1.0, 0.5, 0.25, and 0.1 $\rho_{\rm Jup}$. {\it Bottom panel}: The mass-density diagram of known transiting exoplanets. Dashed lines refer to four planetary models with various core masses and another one without a core \citep{fortney:2007}.}
\label{fig:wasp60_diagrams}
\end{figure}

Fig.~\ref{fig:wasp60_diagrams} shows the change in position in the planet mass-radius diagram (top panel) and  planet mass-density diagram (bottom panel). The revised positions are indicated with green points, while red points indicate the previous values from \citet{hebrard:2013}. The values of the other transiting exoplanets were taken from the TEPCat catalogue\footnote{http://www.astro.keele.ac.uk/jkt/tepcat/}. For illustration, the bottom panel of Fig.~\ref{fig:wasp60_diagrams} shows 1\,Gyr isochrones of exoplanets at 0.045\,au orbital separation from a solar analogue \citep{fortney:2007}, suggesting that the mass of the core of WASP-60\,b should be extremely small.

%

\subsection{Timescales of tidal evolution}
\label{sec:tidal}
All the systems considered in this investigation are likely not synchronised and their total angular momentum is between 0.52 and 0.63 of the minimum critical angular momentum that allows a binary system to reach a stable equilibrium during its tidal evolution \citep{Hut80,Ogilvie14,DamianiLanza15}. Therefore, all our systems are tidally unstable. However, the estimated timescales of tidal evolution are longer than the main-sequence lifetimes of all the systems except for HAT-P-22. Specifically, we can parameterise the efficiency of the dissipation of the tidal kinetic energy inside a star by means of the so-called modified tidal quality factor $Q^{\prime}_{\rm s}$ with a faster dissipation and stronger tidal interaction corresponding to a smaller value of $Q^{\prime}_{\rm s}$ (e.g. \citealp{Ogilvie14}). The value of $Q^{\prime}_{\rm s}$ in exoplanet hosts is very uncertain because of our lack of knowledge of the processes that dissipate the energy of the dynamical tides inside main-sequence stars, but the minimum value of $Q^{\prime}_{\rm s}$ is generally considered to range between $10^{6}$ and $10^{7}$ \citep{OgilvieLin07,Jacksonetal09}. Adopting such a range of values for HAT-P-3, HAT-P-12,WASP-39, and WASP-60, we find orbital decay timescales between a few times and several tens of times of their main-sequence lifetimes by applying a constant $Q^{\prime}_{\rm s}$ version of the tidal evolution model of \citet{Leconteetal10}. The timescale for the evolution of stellar rotation and orbital obliquity is longer than $\sim 20$ Gyr for all these stars. Therefore, their stellar rotation and obliquity have not been significantly affected by tides during their main-sequence evolution. The situation is different in the case of HAT-P-22, which has characteristic timescales of tidal evolution of the rotation and obliquity that range between 0.7 and 7 Gyr for $10^{6} \leq Q^{\prime}_{\rm s} \leq 10^{7}$. In this system, tides may have played a role in reducing an initial obliquity of the orbit and accelerating stellar rotation during the main-sequence lifetime of the star. 

\section{Summary and discussion}
\label{sec:discussion}
\begin{figure*}
\centering
\includegraphics[width=18cm]{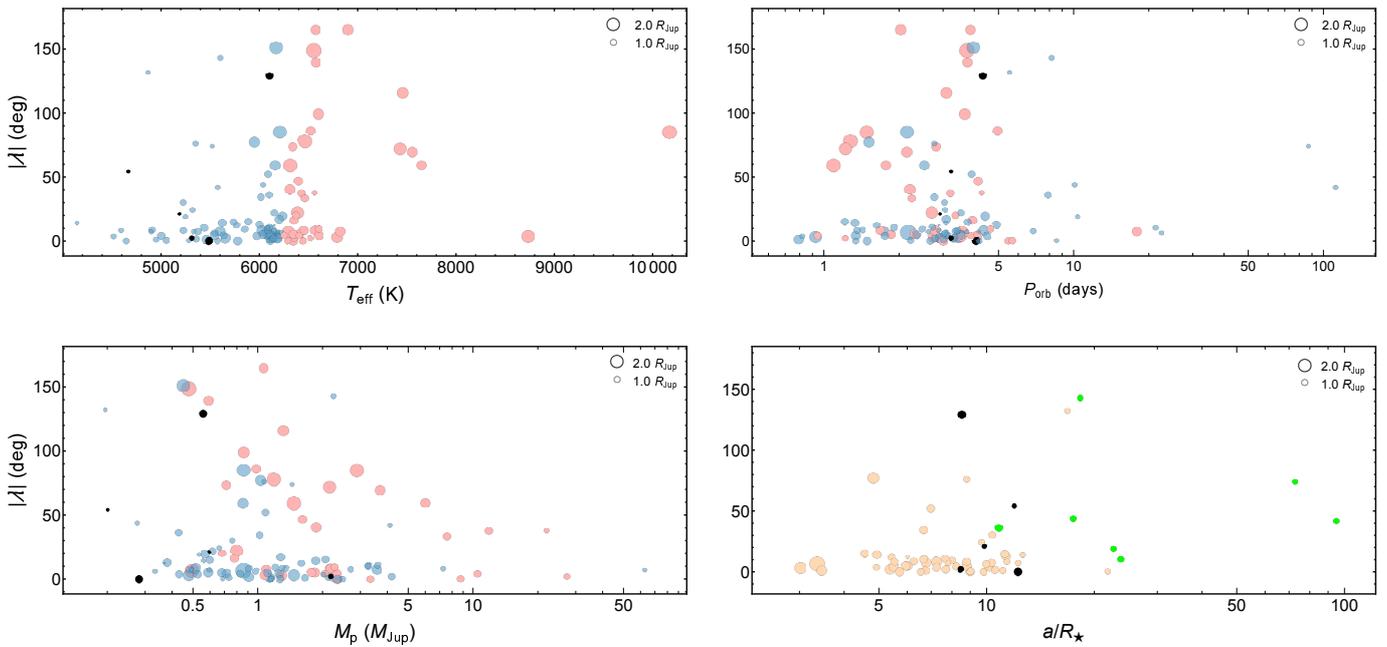}
\caption{Sky-projected orbital obliquity as a function of several stellar, orbital, and planetary parameters. We only considered exoplanets with $R_{\rm p}>0.8\,R_{\rm Jup}$. The data are taken from TEPCat with the exception of the the black points that indicate the five planetary systems examined in this work. {\it Top left panel}: The value $\lambda$ vs. the effective temperature of the parent stars. Blue (red) circles indicate systems in which the parent star has an effective temperature lower (higher) than 6250\,K. The size of each circle is proportional to the corresponding planetary radius. The error bars are suppressed for clarity. {\it Top right panel}: The value $\lambda$ vs. the orbital period of the planets. The symbols are the same as in top left panel. {\it Bottom left panel}: The value $\lambda$ vs. the planet mass. The symbols are the same as in top left panel. {\it Bottom right panel}:\ The value $\lambda$ vs. the scaled orbital distance, $a/R_{\star}$, for those systems with $T_{\rm eff}<6150$\,K \citep{anderson:2015}. The black and light orange circles represent planets with near-circular orbits ($e<0.1$ or consistent with zero; \citealp{bonomo:2017}), while the green circles depict eccentric orbits ($e \geq 0.1$).}
\label{fig:RM_plots}
\end{figure*}

As discussed in Sect.~\ref{sec:introduction}, it is a difficult task to trace all the steps that occurred during the migration phase of hot Jupiters, and the orbital obliquity could play an important role in disentangling the bundle of possible physical processes. For this purpose a statistical study of a large sample of precise measurements of $\lambda$ is mandatory. With the present work, we have given another contribution to the enlargement of the sample, by presenting the first measurements of $\lambda$ for five exoplanetary systems. These systems are HAT-P-3, HAT-P-12, HAT-P-22, WASP-39, and WASP-60, all composed of relatively cold stars, i.e. with $4650$\,K$ < T_{\rm eff} < 5490$\,K, with the exception of WASP-60 for which, based on HARPS-N data, we estimated a hotter temperature, i.e. $T_{\rm eff}=6105 \pm 50$\,K. As shown in Sect.~\ref{sec:alignment}, the spins of two of these systems (HAT-P-22 and WASP-39) are well aligned with the planetary orbital axis. Instead, the orbit of the exoplanet HAT-P-3\,b is slightly misaligned, while that of HAT-P-12\,b seems strongly misaligned, even thought we were not able to put robust constraints on our measurements. Finally, WASP-60\,b is in a retrograde orbit.

All these new values are reported in Fig.~\ref{fig:RM_plots} (black circles), together with the other 107 known transiting exoplanets, for which precise measurements of $\lambda$ exist\footnote{Data taken from TEPCat \citep{southworth:2011}.} and having a radius larger than $0.8\,R_{\rm Jup}$. The last choice was adopted to select a homogeneous sample of giant planets. Basically, we excluded the very few exoplanets in the Neptunian and super-Earth range for which a measurement of the RM effect was performed. The values of $\lambda$ are plotted as a function of stellar effective temperature (top left panel), planetary orbital period (top right panel), planet mass (bottom left panel), and orbital distance in units of stellar radii, $a/R_{\star}$ (bottom right panel). Following \citet{esposito:2017} and references therein, in the first three panels, we divided the data into two groups, according to the temperature of the parent stars. These groups are shown with red circles for $T_{\rm eff} \geq 6250$\,K and blue circles for $T_{\rm eff} < 6250$\,K.

An inspection of the top left panel tells us that the orbit of most of the planets is well aligned for a large range of $T_{\rm eff}$. The region between 6000\,K and 6500\,K appears very crucial, as many planets present a large misalignment. However the presence of several exceptions at lower and higher temperatures make us think there might be an observational bias, implying the necessity to investigate these two zones of the parameter space, where the measurement of $\lambda$ is more challenging.

Whereas the top right panel does not provide any particular insight, the bottom left panel confirms something that was already observed by \citealt{hebrard:2011}, that is the more massive planets ($M_{\rm p} > 4 M_{\rm Jup}$) can be misaligned but are not retrograde.

Finally, in the bottom right panel, we plotted the projected orbital obliquity of the cool-star systems, i.e. following \citet{anderson:2015}, those with $T_{\rm eff} < 6150$\,K., as a function of $a/R_{\star}$. The orange circles represent exoplanets with near-circular orbits ($e < 0.1$) and the green circles eccentric orbits ($e \geq 0.1$). The systems HAT-P-3, HAT-P-12, HAT-P-22, WASP-39, and WASP-60 are again represented with black circles and have $e$ consistent with zero \citep{bonomo:2017}. \citet{anderson:2015} noted that orbits with $a/R_{\star} < 15$ are circular and the corresponding values of $\lambda$ are confined within $\sim 20$ degree of aligned; the distribution of $\lambda$ is broad at greater separations, where eccentric orbits seem to be preferred. Based on a larger sample, the confinement of $\lambda$ is not completely convincing. We stress that any inference on the possible trend of higher eccentricity, larger separation companions that are found on typically misaligned orbits, still relies on small-number statistics.

To summarise the results of the photometric time series, in two cases (WASP-60 and  HAT-P-12) there is no clear periodicity. In one case (WASP-39) we found a possible periodicity, but it is probably spurious. In another case (HAT-P-3) we detected a periodicity that is promising but below the level of significance. Finally, in the fifth case (HAT-P-22) we were able to propose a more robust detection of the rotational period, i.e.\ $28.7 \pm 0.4$~days.
Knowing the rotation period of HAT-P-22, we can measure its rotational velocity and, therefore, the inclination of its spin. It is then straightforward to estimate the true misalignment angle of HAT-P-22\,b, which is $\psi=25^{\circ}\pm 18^{\circ}$. A more precise determination of the $v\sin{i_\star}$ is necessary to better constrain this value.

We finally reviewed the main physical parameters of the five exoplanetary systems. This analysis was performed following the {\it homogeneous studies approach} \citep{southworth:2012}, and based on the HARPS-N data and on both 17 new photometric light curves and others taken from the literature. In four cases (see Tables \ref{tab:hatp3_final_parameters}, \ref{tab:hatp12_final_parameters}, \ref{tab:hatp22_final_parameters}, and \ref{tab:wasp39_final_parameters} in Appendix~\ref{appendix_PhysicalParameters}), we found good agreements with the values reported in the literature. However, for the case of the WASP-60 planetary system, many of the physical characteristics of both the planet and star are very different from those reported by \citet{hebrard:2013}; see Table~\ref{tab:wasp60_final_parameters}. In particular WASP-60\,b is larger and therefore much less dense.
We emphasise that, based on our photometric follow-up monitoring of a complete transit of WASP-60\,b, we measured a transit depth $\approx 33\%$ deeper than that presented in the discovery paper, which is only based on survey data and on a single incomplete follow-up light curve \citep{hebrard:2013}. The modelling of incomplete transit light curves can lead to incorrect estimation of the photometric parameters ($R_{\rm p}/R_{\star}$, $a/R_{\star}$, i), contact points and, therefore, stellar density \citep{mancini:2016}. The present case of WASP-60 is a good example of this effect.

\begin{acknowledgements}
This paper is based on observations collected with the following telescopes: the 3.58\,m Telescopio Nazionale Galileo (TNG), operated on the island of La Palma (Spain) by the Fundaci\'{o}n Galileo Galilei of the INAF (Istituto Nazionale di Astrofisica) at the Spanish Observatorio del Roque de los Muchachos of the Instituto de Astrof\'{i}sica de Canarias, in the frame of the programme Global Architecture of Planetary Systems (GAPS); the CAHA 2.2.\,m and 1.23\,m telescopes at the Centro Astron\'{o}mico Hispano Alem\'{a}n (CAHA) in Calar Alto (Spain); the Cassini 1.52\,m telescope at the Astronomical Observatory of Bologna in Loiano (Italy); the Isaac Newton Telescope, operated on the island of La Palma, by the Isaac Newton Group in the Spanish Observatorio del Roque de los Muchachos. The HARPS-N instrument has been built by the HARPS-N Consortium, a collaboration between the Geneva Observatory (PI Institute), the Harvard-Smithonian Center for Astrophysics, the University of St. Andrews, the University of Edinburgh, the Queen's University of Belfast, and INAF. Operations at the Calar Alto telescopes are jointly performed by the Max-Planck Institut f\"{u}r Astronomie (MPIA) and the Instituto de Astrof\'{i}sica de Andaluc\'{i}a (CSIC). The reduced light curves presented in this work will be made available at the CDS (http://cdsweb.u-strasbg.fr/). The GAPS project in Italy acknowledges support from INAF through the ``Progetti Premiali'' funding scheme of the Italian Ministry of Education, University, and Research. 
We thank Roberto Gualandi for his technical assistance at the Cassini telescope.
We thank the support astronomers of CAHA for their technical assistance at the Zeiss telescope.
The research leading to these results has received funding from the European Union Seventh Framework Programme (FP7/2007-2013) under Grant Agreement No. 313014 (ETAEARTH).
D.F.E. acknowledges funding from the Science and Technology Facilities Council in the form of a studentship.
We acknowledge the use of the following internet-based resources: the ESO Digitized Sky Survey;
the TEPCat catalogue; the SIMBAD data base operated at CDS, Strasbourg, France; and the arXiv scientific paper preprint
service operated by Cornell University.
\end{acknowledgements}


\bibliographystyle{aa}


\begin{appendix}

\section{Details of the spectroscopic and photometric observations} 
\label{appendix_logs}
The tables in this appendix report the details of the spectroscopic and photometric observations that were presented in this work.
In particular, Table~\ref{tab:tng-log} shows the log of the HARPS-N observations, while Table~\ref{tab:logs} the log concerning all the photometric follow-up observations.

%
\begin{table*}
{\small
\caption{Details of the spectroscopic HARPS-N observations presented in this work.}
\label{tab:tng-log}
\centering
\begin{tabular}{l c c c c c c c c c}     
\hline\hline \\ [-8pt]
Object & Type & Date\tablefootmark{(a)} & UT Start & UT End & $N_{\rm obs}$ & $T_{\rm exp}$[s] & Airmass\tablefootmark{(b)} & Moon\tablefootmark{(c)} & 2$^{\rm nd}$ fibre \\
\hline \\ [-6pt]
HAT-P-3 & K5 & 2013.06.10 & 20:58 &  01:03   & 22    &  600   &  1.07$\rightarrow$1.06$\rightarrow$1.38 & NO &    ThAr lamp \\ [2pt]
\hline \\ [-6pt]
WASP-60 & ~~~~F9\tablefootmark{(d)} & 2013.10.20 & 21:08 & 02:59 & 21 & 900 & 1.07$\rightarrow$1.00$\rightarrow$1.59 & 95\%/49$^{\circ}$ & Sky \\ [2pt]
\hline \\ [-6pt]
HAT-P-22 & G2 & 2014.04.03 &  22:25     &  03:47   & 31    &  600   &  1.06$\rightarrow$1.07$\rightarrow$1.96  &  NO & ThAr lamp \\ [2pt]
\hline \\ [-6pt]
HAT-P-12 & K5 & $\begin{tabular}{c}
2015.03.13\\
2015.04.24 \\
\end{tabular}$ & $\begin{tabular}{c}
00:59 \\
20:56 \\
\end{tabular}$ & $\begin{tabular}{c}
06:22 \\
01:02 \\
\end{tabular}$ & $\begin{tabular}{c}
21(17) \\
16 \\
\end{tabular}$ & $\begin{tabular}{c}
900 \\
900 \\
\end{tabular}$ & $\begin{tabular}{c}
1.20$\rightarrow$1.03$\rightarrow$1.22 \\
1.48$\rightarrow$1.03 \\
\end{tabular}$ & $\begin{tabular}{c}
46\%/81$^{\circ}$ \\
41\%/82$^{\circ}$ \\
\end{tabular}$ & $\begin{tabular}{c}
Sky \\
Sky \\
\end{tabular}$ \\ [2pt] %
\hline \\ [-6pt]
WASP-39 & K5 & 2015.05.04 & 22:54 & 04:02 & 20 & 900 & 1.32$\rightarrow$1.18$\rightarrow$1.79  &  99\%/19$^{\circ}$\tablefootmark{(e)} & Sky \\ [2pt]
\hline
\end{tabular}
\tablefoot{
\tablefoottext{a}{Dates refer to the beginning of the night.}
\tablefoottext{b}{Values at first$\rightarrow$last exposure, or first$\rightarrow$meridian$\rightarrow$last exposure.}
\tablefoottext{c}{Fraction of illumination and angular distance from the target.}
\tablefoottext{d}{This work.}
\tablefoottext{e}{We checked that the nearby full Moon (${\rm RV}=-3.5$\,km\,s$^{-1}$) did not contaminate the CCF profiles and had no effect on the RV measurements (see \citealt{esposito:2017}).}
}
}
\end{table*}
\begin{table*}
\caption{Details of the photometric follow-up observations presented in this work.} %
\label{tab:logs} %
\centering     %
\tiny          %
\setlength{\tabcolsep}{5pt}
\begin{tabular}{lcccccccccc}
\hline\hline\\[-6pt]
Telescope & Date of   & Start time & End time  &$N_{\rm obs}$ & $T_{\rm exp}$ & Filter & Airmass & Moon & Aperture   & Scatter \\
          & first obs &    (UT)    &   (UT)    &              & (s)                     &        &         &illum.& radii (px) & (mmag)  \\
\hline \\[-6pt]
\multicolumn{10}{l}{\textbf{HAT-P-3:}} \\
Cassini     & 2010.06.07 & 20:12 & 00:16 & 107 & 60-150  & Gunn    $i$ & $1.01 \rightarrow 1.35$ &  ~~22\% & 16,\,26,\,50 & 1.69 \\
CA\,1.23\,m & 2015.05.01 & 20:09 & 03:52 & 246 & 85      & Cousins $I$ & $1.22 \rightarrow 1.03 \rightarrow 1.50$ &  ~~96\% & 17,\,40,\,60 & 1.49 \\
CA\,1.23\,m & 2015.05.07 & 19:45 & 02:23 & 178 & 102-135 & Cousins $I$ & $1.21 \rightarrow 1.03 \rightarrow 1.28$ &  ~~84\% & 22,\,45,\,70 & 0.66 \\
CA\,1.23\,m & 2015.06.02 & 20:28 & 03:01 & 183 & 98-130 & Cousins $I$ & $1.04 \rightarrow 1.03 \rightarrow 2.20$ &  100\% & 20,\,45,\,70 & 0.83 \\ [6pt] %
\multicolumn{10}{l}{\textbf{HAT-P-12:}} \\
INT   & 2010.04.29 & 02:28 & 05:47 & ~~91 & 100  & Gunn    $r$ & $1.12 \rightarrow 2.15$ &  ~~97\% & 20,\,40,\,60 & 0.92 \\  %
CA\,2.2\,m & 2010.05.11 & 21:11 & 03:54 & 151 & 80-120 & Str\"{o}mgren $u$ & $1.07 \rightarrow 1.03 \rightarrow 1.74$ &  ~~~~5\% & ~~9,\,14,\,30 & 5.97 \\
CA\,2.2\,m & 2010.05.11 & 21:11 & 03:54 & 157 & 80-120 & Gunn $g$ & $1.07 \rightarrow 1.03 \rightarrow 1.74$ &  ~~~~5\% & 25,\,35,\,60 & 1.14 \\
CA\,2.2\,m & 2010.05.11 & 21:11 & 03:54 & 146 & 80-120 & Gunn $r$ & $1.07 \rightarrow 1.03 \rightarrow 1.74$ &  ~~~~5\% & 28,\,38,\,70 & 1.04 \\
CA\,2.2\,m & 2010.05.11 & 21:11 & 03:54 & 166 & 80-120 & Johnson $I$ & $1.07 \rightarrow 1.03 \rightarrow 1.74$ &  ~~~~5\% & 25,\,35,\,60 & 1.32 \\
CA\,1.23\,m & 2012.03.06 & 23:41 & 05:33 & ~~83 & 105-130 & Cousins $R$ & $1.31 \rightarrow 1.00 \rightarrow 1.13$ &  ~~97\% & 24,\,45,\,70 & 1.38 \\
Cassini     & 2012.04.21 & 01:38 & 03:24 & ~~47 & 120  & Gunn    $r$ & $1.20 \rightarrow 1.61$ &  ~~~~0\% & 20,\,55,\,85 & 1.27 \\ %
CA\,1.23\,m & 2013.06.15 & 20:48 & 01:44 & ~146 & 70-120 & Cousins $R$ & $1.00 \rightarrow 2.05$ &  ~~41\% & 23,\,48,\,70 & 1.22 \\
CA\,1.23\,m & 2014.03.16 & 00:55 & 05:23 & ~152 & 85-100 & Cousins $I$ & $1.07 \rightarrow 1.00 \rightarrow 1.17$ &  ~100\% & 28,\,50,\,72 & 1.06 \\
CA\,1.23\,m & 2014.04.13 & 20:58 & 03:42 & ~156 & 120-150 & Cousins $I$ & $1.44 \rightarrow 1.00 \rightarrow 1.21$ &  ~~98\% & 35,\,55,\,75 & 0.94 \\
CA\,1.23\,m & 2015.06.24 & 21:58 & 01:35 & ~~95 & 120 & Cousins $R$ & $1.07 \rightarrow 2.13$ &  ~~55\% & 26,\,54,\,80 & 1.49 \\
CA\,1.23\,m & 2016.07.04 & 20:33 & 01:29 & ~164 & 95 & Cousins $R$ & $1.04 \rightarrow 2.66$ &  ~~~~1\% & 18,\,35,\,50 & 1.62 \\  [6pt]
\multicolumn{10}{l}{\textbf{WASP-60:}} \\
CA\,12.3\,m & 2014.10.21 & 18:40 & 01:57 & 233 & 100 & Cousins $I$ & $1.70 \rightarrow 1.00 \rightarrow 1.28$ &  ~~~~3\% & 19,\,27,\,50 & 0.68 \\ [6pt] %
\hline
\end{tabular}
\tablefoot{$N_{\rm obs}$ is the number of observations, $T_{\rm
exp}$ is the exposure time, and `Moon illum.' is the geocentric fractional
illumination of the Moon at midnight (UT). The aperture sizes are
the radii of the software apertures for the star, inner sky, and
outer sky, respectively. Scatter is the \emph{rms} scatter of the
data vs. a fitted model.}
\end{table*}

\section{HARPS-N RV measurements} 
\label{appendix_RV}

The RV measurements, which were obtained with HARPS-N (this work), are reported in this appendix.

\begin{table}
\caption{HARPS-N RV measurements of HAT-P-3.} %
\label{tab:RV_HAT-P-3} %
\centering     %
\tiny          %
\setlength{\tabcolsep}{8pt}
\begin{tabular}{cccc}
\hline\hline\\ [-6pt]
BJD(TDB) & RV\,(m\,s$^{-1}$) & Error\,(m\,s$^{-1}$) & S/N  \\
\hline\\ [-6pt]
2\,456\,454.378444  & $-23367.2$ &  2.0 & 39.1 \\
2\,456\,454.385689  & $-23371.8$ &  2.1 & 38.9 \\
2\,456\,454.392929  & $-23372.4$ &  2.1 & 39.6 \\
2\,456\,454.400166  & $-23373.7$ &  1.9 & 41.6 \\
2\,456\,454.407406  & $-23373.5$ &  1.9 & 42.2 \\
2\,456\,454.414647  & $-23377.3$ &  1.9 & 42.0 \\
2\,456\,454.421896  & $-23375.2$ &  1.9 & 41.8 \\
2\,456\,454.429142  & $-23374.5$ &  1.0 & 40.7 \\
2\,456\,454.436382  & $-23378.5$ &  2.2 & 38.2 \\
2\,456\,454.443623  & $-23380.5$ &  2.6 & 33.5 \\
2\,456\,454.450868  & $-23384.2$ &  4.2 & 23.0 \\
2\,456\,454.458118  & $-23389.8$ &  3.1 & 28.8 \\
2\,456\,454.465363  & $-23393.0$ &  2.7 & 32.5 \\
2\,456\,454.481173  & $-23398.0$ &  2.7 & 32.3 \\
2\,456\,454.488422  & $-23403.4$ &  3.6 & 25.9 \\
2\,456\,454.495667  & $-23395.2$ &  3.9 & 24.7 \\
2\,456\,454.502908  & $-23394.1$ &  5.0 & 20.5 \\
2\,456\,454.512501  & $-23402.5$ &  4.5 & 22.2 \\
2\,456\,454.519746  & $-23400.1$ &  7.0 & 15.8 \\
2\,456\,454.526986  & $-23400.9$ &  5.3 & 19.6 \\
2\,456\,454.534227  & $-23399.6$ &  6.1 & 17.5 \\
2\,456\,454.541477  & $-23397.2$ &  4.5 & 22.3 \\ [2pt] %
\hline
\end{tabular}
\end{table}

\begin{table}
\caption{HARPS-N RV measurements of HAT-P-12.} %
\label{tab:RV_HAT-P-12} %
\centering     %
\tiny          %
\setlength{\tabcolsep}{8pt}
\begin{tabular}{cccc}
\hline\hline\\ [-6pt]
BJD(TDB) & RV\,(m\,s$^{-1}$) & Error\,(m\,s$^{-1}$) & S/N  \\
\hline\\ [-6pt]
2\,457\,095.593419 & $-40450.2$ &  6.0 & 17.0 \\
2\,457\,095.604117 & $-40452.4$ &  6.6 & 15.9 \\
2\,457\,095.614820 & $-40456.4$ &  7.6 & 14.6 \\
2\,457\,095.625522 & $-40455.5$ & 11.1 & 10.8 \\
2\,457\,095.636220 & $-40451.7$ &  6.9 & 15.6 \\
2\,457\,095.646928 & $-40448.7$ &  5.9 & 17.4 \\
2\,457\,095.657626 & $-40459.3$ &  6.1 & 17.1 \\
2\,457\,095.668328 & $-40456.7$ &  7.9 & 14.3 \\
2\,457\,095.679040 & $-40464.1$ &  6.8 & 15.7 \\
2\,457\,095.689748 & $-40466.1$ &  7.3 & 15.0 \\
2\,457\,095.700450 & $-40463.9$ &  5.6 & 18.1 \\
2\,457\,095.711153 & $-40465.3$ &  8.4 & 13.3 \\
2\,457\,095.721860 & $-40454.8$ & 10.5 & 11.4 \\
2\,457\,095.732559 & $-40467.9$ & 11.9 & 10.3 \\
2\,457\,095.743267 & $-40471.2$ &  8.6 & 13.4 \\
2\,457\,095.753974 & $-40466.5$ &  8.3 & 13.9 \\
2\,457\,095.764682 & $-40466.7$ &  6.7 & 16.3 \\
2\,457\,137.381494 & $-40462.4$ & 10.2 & 12.3 \\
2\,457\,137.392206 & $-40453.1$ &  9.5 & 13.0 \\
2\,457\,137.402913 & $-40462.6$ &  8.6 & 14.0 \\
2\,457\,137.413619 & $-40436.0$ &  8.6 & 14.2 \\
2\,457\,137.424331 & $-40440.6$ &  7.8 & 15.2 \\
2\,457\,137.435038 & $-40448.0$ &  7.0 & 16.2 \\
2\,457\,137.445740 & $-40454.1$ &  7.0 & 16.0 \\
2\,457\,137.456452 & $-40461.4$ &  9.6 & 13.1 \\
2\,457\,137.467149 & $-40457.3$ &  8.6 & 14.1 \\
2\,457\,137.477861 & $-40476.8$ &  8.5 & 14.1 \\
2\,457\,137.488563 & $-40457.1$ &  7.3 & 15.5 \\
2\,457\,137.499261 & $-40466.6$ &  8.4 & 14.4 \\
2\,457\,137.509963 & $-40460.5$ &  7.8 & 15.1 \\
2\,457\,137.520678 & $-40469.5$ &  6.2 & 17.4 \\
2\,457\,137.531385 & $-40463.8$ &  5.2 & 19.9 \\
2\,457\,137.542087 & $-40467.9$ &  5.6 & 18.7 \\[2pt] %
\hline
\end{tabular}
\end{table}

\begin{table}
\caption{HARPS-N RV measurements of HAT-P-22.} %
\label{tab:RV_HAT-P-22} %
\centering     %
\tiny          %
\setlength{\tabcolsep}{8pt}
\begin{tabular}{cccc}
\hline\hline\\ [-6pt]
BJD(TDB) & RV\,(m\,s$^{-1}$) & Error\,(m\,s$^{-1}$) & S/N  \\
\hline\\ [-6pt]
2\,456\,751.440813 & 12706.6 & 1.0 & 75.6 \\
2\,456\,751.448034 & 12704.0 & 1.0 & 74.1 \\
2\,456\,751.455255 & 12698.5 & 0.9 & 78.5 \\
2\,456\,751.462476 & 12695.3 & 0.9 & 77.7 \\
2\,456\,751.469698 & 12692.3 & 0.9 & 79.1 \\
2\,456\,751.476918 & 12686.9 & 0.9 & 79.0 \\
2\,456\,751.484144 & 12682.6 & 0.9 & 76.5 \\
2\,456\,751.491369 & 12679.0 & 0.9 & 76.5 \\
2\,456\,751.498599 & 12674.0 & 0.9 & 77.4 \\
2\,456\,751.505829 & 12673.3 & 1.0 & 75.1 \\
2\,456\,751.513055 & 12674.7 & 1.2 & 64.4 \\
2\,456\,751.520281 & 12671.4 & 1.4 & 54.4 \\
2\,456\,751.527507 & 12663.7 & 1.4 & 56.1 \\
2\,456\,751.534733 & 12661.5 & 1.5 & 53.2 \\
2\,456\,751.541962 & 12649.6 & 1.5 & 52.9 \\
2\,456\,751.549192 & 12645.5 & 1.6 & 49.3 \\
2\,456\,751.556422 & 12638.4 & 1.5 & 50.8 \\
2\,456\,751.563635 & 12630.0 & 1.4 & 54.1 \\
2\,456\,751.570857 & 12624.3 & 1.4 & 55.1 \\
2\,456\,751.578086 & 12616.2 & 1.7 & 47.5 \\
2\,456\,751.585320 & 12606.8 & 2.2 & 38.5 \\
2\,456\,751.592554 & 12604.2 & 1.8 & 44.1 \\
2\,456\,751.599775 & 12595.6 & 1.7 & 47.4 \\
2\,456\,751.606998 & 12602.5 & 1.6 & 48.7 \\
2\,456\,751.614219 & 12600.7 & 1.4 & 57.0 \\
2\,456\,751.621450 & 12595.8 & 1.4 & 54.2 \\
2\,456\,751.628670 & 12591.0 & 1.4 & 56.9 \\
2\,456\,751.635900 & 12586.9 & 1.3 & 59.6 \\
2\,456\,751.643130 & 12583.6 & 1.4 & 57.1 \\
2\,456\,751.650352 & 12579.3 & 1.4 & 57.6 \\
2\,456\,751.657573 & 12573.8 & 1.5 & 53.7 \\ [2pt] %
\hline
\end{tabular}
\end{table}

\begin{table}
\caption{HARPS-N RV measurements of WASP-39.} %
\label{tab:RV_WASP-39} %
\centering     %
\tiny          %
\setlength{\tabcolsep}{8pt}
\begin{tabular}{cccc}
\hline\hline\\ [-6pt]
BJD(TDB) & RV\,(m\,s$^{-1}$) & Error\,(m\,s$^{-1}$) & S/N  \\
\hline\\ [-6pt]
2\,457\,147.465780 & $-58445.5$ & 9.7 & 14.2 \\
2\,457\,147.476477 & $-58438.3$ & 9.6 & 14.1 \\
2\,457\,147.487180 & $-58439.0$ & 6.2 & 19.7 \\
2\,457\,147.497878 & $-58431.0$ & 5.7 & 21.2 \\
2\,457\,147.508575 & $-58444.0$ & 7.5 & 17.3 \\
2\,457\,147.519282 & $-58436.0$ & 4.9 & 23.8 \\
2\,457\,147.529990 & $-58417.3$ & 4.7 & 24.4 \\
2\,457\,147.540692 & $-58434.3$ & 5.1 & 22.8 \\
2\,457\,147.551390 & $-58435.3$ & 5.2 & 22.6 \\
2\,457\,147.562097 & $-58436.9$ & 5.0 & 23.2 \\
2\,457\,147.572794 & $-58437.5$ & 5.0 & 23.5 \\
2\,457\,147.583501 & $-58448.9$ & 5.0 & 23.4 \\
2\,457\,147.594208 & $-58453.1$ & 4.5 & 25.0 \\
2\,457\,147.604906 & $-58466.1$ & 4.7 & 24.4 \\
2\,457\,147.615604 & $-58459.3$ & 4.5 & 25.3 \\
2\,457\,147.626324 & $-58448.4$ & 4.8 & 24.3 \\
2\,457\,147.637027 & $-58446.9$ & 4.7 & 24.6 \\
2\,457\,147.647729 & $-58432.9$ & 5.3 & 23.1 \\
2\,457\,147.658427 & $-58452.3$ & 4.6 & 25.3 \\
2\,457\,147.669133 & $-58445.1$ & 5.9 & 20.9 \\ [2pt] %
\hline
\end{tabular}
\end{table}

\begin{table}
\caption{HARPS-N RV measurements of WASP-60.} %
\label{tab:RV_WASP-60} %
\centering     %
\tiny          %
\setlength{\tabcolsep}{8pt}
\begin{tabular}{cccc}
\hline\hline\\ [-6pt]
BJD(TDB) & RV\,(m\,s$^{-1}$) & Error\,(m\,s$^{-1}$) & S/N \\
\hline\\ [-6pt]
2\,456\,586.391049 & $-26529.3$ & 6.0 &  25.4 \\
2\,456\,586.409541 & $-26520.0$ & 5.6 &  27.0  \\
2\,456\,586.420258 & $-26530.0$ & 5.3 &  27.8  \\
2\,456\,586.430971 & $-26527.3$ & 5.8 &  26.1  \\
2\,456\,586.441693 & $-26516.7$ & 6.2 &  24.5  \\
2\,456\,586.452402 & $-26533.3$ & 6.1 &  24.9  \\
2\,456\,586.463115 & $-26524.3$ & 5.9 &  25.1  \\
2\,456\,586.473828 & $-26521.7$ & 6.5 &  23.5  \\
2\,456\,586.494129 & $-26516.8$ & 5.4 &  28.0  \\
2\,456\,586.504856 & $-26525.2$ & 6.2 &  25.3  \\
2\,456\,586.515582 & $-26515.0$ & 5.0 &  29.4  \\
2\,456\,586.526300 & $-26512.1$ & 5.2 &  28.7  \\
2\,456\,586.537013 & $-26524.9$ & 6.4 &  24.6  \\
2\,456\,586.547726 & $-26516.0$ & 6.9 &  23.5  \\
2\,456\,586.558439 & $-26517.0$ & 5.5 &  27.9  \\
2\,456\,586.569156 & $-26533.6$ & 5.7 &  27.3  \\
2\,456\,586.579878 & $-26536.4$ & 5.5 &  28.4  \\
2\,456\,586.590596 & $-26532.6$ & 6.3 &  25.8  \\
2\,456\,586.601309 & $-26541.0$ & 6.1 &  26.6  \\
2\,456\,586.612026 & $-26544.7$ & 6.1 &  26.5  \\
2\,456\,586.624506 & $-26539.0$ & 5.7 &  28.1  \\ [2pt] %
\hline
\end{tabular}
\end{table}
%

\section{Revised physical parameters of the five planetary systems} 
\label{appendix_PhysicalParameters}

The tables in this appendix report the values of the main physical parameters of the five planetary systems under study. The values obtained in this work (Sect.~\ref{sec:physical_parameters}) are compared with those taken from the literature. Where two error bars are given, the first refers to the statistical uncertainties and the second to the systematic errors. \\ %

\begin{table*}
\tiny
\centering %
\caption{Physical parameters of the planetary system HAT-P-3 derived in this work.} %
\label{tab:hatp3_final_parameters} %
\begin{tabular}{l c c c c c}
\hline %
\hline  \\[-8pt]
Parameter & Nomen. & Unit & This Work & \citet{torres:2008} & \citet{southworth:2012} \\
\hline  \\[-6pt]
\multicolumn{1}{l}{\textbf{Stellar parameters}} \\ [2pt] %
Effective temperature         \dotfill & $T_{\mathrm{eff}}$   & K            & $5190  \pm 80  $ & $5185  \pm 80$   & -- \\ [2pt] %
Iron abundance               \dotfill & [Fe/H]               &              & $+0.24 \pm 0.08$ & $+0.27 \pm 0.00$ & -- \\ [2pt] %
Projected rotational velocity \dotfill & $v\,\sin{i_{\star}}$ & km\,s$^{-1}$ & $1.4   \pm 0.5 $ & $0.5   \pm 0.5$  & -- \\ [2pt] %
Mass    \dotfill & $M_{\star}$ & $M_{\sun}$ & $0.925 \pm 0.031 \pm 0.034$ & $0.928_{-0.054}^{+0.044}$ & $0.900 \pm 0.036 \pm 0.044$ \\ [2pt]%
Radius  \dotfill & $R_{\star}$ & $R_{\sun}$ & $0.850 \pm 0.021 \pm 0.010$ & $0.833_{-0.044}^{+0.034}$ & $0.870 \pm 0.016 \pm 0.014$ \\ [2pt] %
Mean density \dotfill & $\rho_{\star}$ & $\rho_{\sun}$ & $1.51 \pm 0.11 $ & $1.90_{-0.42}^{+0.38}$    & $1.365 \pm 0.078$ \\ [2pt] %
Logarithmic surface gravity \dotfill & $\log{g_{\star}}$ & cgs & $4.545\pm0.022\pm0.005$ & $4.564\pm0.032$&$4.513\pm0.020\pm 0.007$ \\ [2pt]%
Age \dotfill & & Gyr & $2.9_{-3.7\,-3.2}^{+1.7\,+2.1} $ & $1.5_{-1.4}^{+5.4}$ & $7.5_{-3.8\,-2.7}^{+4.2\,+3.6} $\\ [2pt]  %
\hline \\[-6pt]%
\multicolumn{1}{l}{\textbf{Planetary parameters}} \\ [2pt] %
Mass \dotfill & $M_{\mathrm{p}}$ & $M_{\mathrm{Jup}}$ & $0.595\pm0.019\pm0.015$&$0.596_{-0.026}^{+0.024}$&$0.584\pm0.020\pm0.019$ \\ [2pt] %
Radius \dotfill & $R_{\mathrm{p}}$ & $R_{\mathrm{Jup}}$&$0.911\pm0.032\pm0.011$&$0.899_{-0.049}^{+0.043}$&$0.947\pm0.027\pm 0.015$ \\ [2pt] %
Mean density \dotfill&$\rho_{\mathrm{p}}$&$\rho_{\mathrm{Jup}}$&$0.735\pm0.075\pm0.009$&$0.77_{-011}^{+0.14}$&$0.643\pm0.052\pm 0.011$\\[2pt]%
Surface gravity  \dotfill & $g_{\mathrm{p}}$ & m\,s$^{-2}$ & $17.8 \pm 1.2$ & $20.4_{-3.1}^{+3.0}$ & $16.14 \pm 0.90$ \\ [2pt] %
Equilibrium temperature \dotfill & $T_{\mathrm{eq}}$ & K & $1170 \pm 17$ & $1127_{-39}^{+49}$ & $1189 \pm 16$ \\  [2pt] %
Safronov number \dotfill & $\Theta$ & & $0.0547 \pm 0.0022 \pm 0.0007$ & $0.0585_{-0.0048}^{+0.0044}$ & $0.0526\pm 0.0019\pm 0.0009$ \\ [2pt]%
\hline \\[-6pt]%
\multicolumn{1}{l}{\textbf{Orbital parameters}} \\ [2pt] %
Time of mid-transit \dotfill & $T_{0}$            & BJD(TDB)& $2\,457\,150.39472\,(58)$ & $2\,454\,218.7594\,(29) $ & -- \\ [2pt] %
Period              \dotfill & $P_{\mathrm{orb}}$ & days    & $2.89973838\,(27)$ & $2.899703\,(54) $ & 2.8997360\,(20) \\  [2pt]%
Semi-major axis \dotfill & $a$ & au & $0.03878\pm0.00044\pm0.00048$ & $0.03882_{-0.00077}^{+0.00060}$&$0.03842\pm0.00050\pm 0.00063$ \\ [2pt]%
Inclination \dotfill & $i$ & degree & $86.31 \pm 0.19$ & $87.24 \pm 0.69$ & $86.15 \pm 0.19$ \\ %
RV-curve semi-amplitude \dotfill & $K_{\rm A}$ &  m\,s$^{-1}$ & $90.63 \pm 0.58\tablefootmark{a}$ & $89.1 \pm 2.0$ & -- \\ [2pt] %
Barycentric RV  \dotfill & $\gamma$ & km\,s$^{-1}$ & $-23.3849\pm 0.0007  $   & $-14.8 \pm 0.10$  & --\\  [2pt] %
Projected spin-orbit angle  \dotfill & $\lambda$ & degree  & $21.2 \pm 8.7$ & -- & -- \\ [2pt] %
\hline %
\end{tabular}
\tablefoot{
\tablefoottext{a}{This value of $K_{\mathrm{A}}$ was determined from out-of-transit RV HIRES+HARPS-N data.}}
\end{table*}

\begin{table*}
\tiny
\centering %
\caption{Physical parameters of the planetary system HAT-P-12 derived in this work.} %
\label{tab:hatp12_final_parameters} %
\begin{tabular}{l c c c c c}
\hline %
\hline  \\[-8pt]
Parameter & Nomen. & Unit & This Work & \citet{hartman:2009} & \citet{lee:2012} \\
\hline  \\[-6pt]
\multicolumn{1}{l}{\textbf{Stellar parameters}} \\[2pt] %
Effective temperature         \dotfill & $T_{\mathrm{eff}}$   & K            & $4665  \pm 45$   & $4650  \pm 60$   & -- \\ [2pt] %
Iron abundance               \dotfill & [Fe/H]               &              & $-0.20 \pm 0.09$ & $-0.29 \pm 0.05$ & -- \\ [2pt] %
Projected rotational velocity \dotfill & $v\,\sin{i_{\star}}$ & km\,s$^{-1}$ & $0.5   \pm 0.5 $ & $0.5   \pm 0.4$  & -- \\ [2pt] %
Mass    \dotfill & $M_{\star}$ & $M_{\sun}$ & $0.691 \pm 0.032 \pm 0.015$ & $0.733 \pm 0.018$ & $0.727 \pm 0.019 $ \\ [2pt]%
Radius  \dotfill & $R_{\star}$ & $R_{\sun}$ & $0.679 \pm 0.012 \pm 0.005$ & $0.701^{+0.017}_{-0.012}$ & $0.702 \pm 0.013$ \\ [2pt] %
Mean density \dotfill & $\rho_{\star}$ & $\rho_{\sun}$ & $2.205 \pm 0.077$ & $-$ & $2.100 \pm 0.089$ \\ [2pt] %
Logarithmic surface gravity \dotfill & $\log{g_{\star}}$ & cgs & $4.614 \pm 0.012 \pm 0.003$ & $4.75 \pm 0.10$ & $4.607 \pm 0.020$ \\ %
Age \dotfill & & Gyr & $7.2_{-4.4\,-2.8}^{+3.7\,+5.3} $ & $2.5 \pm 2.0$ & $3.2 \pm 3.8$\\ [2pt]  %
\hline \\[-6pt]%
\multicolumn{1}{l}{\textbf{Planetary parameters}} \\ [2pt] %
Mass \dotfill   & $M_{\mathrm{p}}$ & $M_{\mathrm{Jup}}$ & $0.201 \pm 0.011 \pm 0.003$ & $0.211 \pm 0.012$ & $0.210 \pm 0.012$ \\ [2pt] %
Radius \dotfill & $R_{\mathrm{p}}$ & $R_{\mathrm{Jup}}$ & $0.919 \pm 0.022 \pm 0.007$ & $0.959^{+0.029}_{-0.021}$ & $0.936\pm0.012$ \\ [2pt] %
Mean density \dotfill & $\rho_{\mathrm{p}}$ & $\rho_{\mathrm{Jup}}$ & $0.242\pm 0.017\pm0.002$ & $0.222\pm0.019$ & $0.240\pm0.012$ \\ [2pt]  %
Surface gravity  \dotfill & $g_{\mathrm{p}}$ & m\,s$^{-2}$ & $5.89 \pm 0.34$ & $5.6 \pm 0.4 $ & $6.37 \pm 0.30$ \\ [2pt] %
Equilibrium temperature \dotfill & $T_{\mathrm{eq}}$ & K & $955 \pm  11$ & $963 \pm 16 $ & $ 960 \pm 14$ \\  [2pt] %
Safronov number \dotfill & $\Theta$ & & $0.0238 \pm 0.0012 \pm 0.0002$ & $0.023 \pm 0.001$ & $0.0236 \pm 0.0015$ \\  [2pt]%
\hline \\[-6pt]%
\multicolumn{1}{l}{\textbf{Orbital parameters}} \\ [2pt] %
Time of mid-transit \dotfill & $T_{0}$ & BJD(TDB)& $2\,455\,328.49068\,(19)$ & $2\,454\,419.19556\,(20) $ & 2\,454\,187.85560\,(11) \\ [2pt] %
Period              \dotfill & $P_{\mathrm{orb}}$ & days    & $3.21305992\,(35)$ & $3.2130598\,(21)$ & 3.21305961\,(35) \\  [2pt]%
Semi-major axis \dotfill & $a$ & au & $0.03767 \pm 0.00057 \pm 0.00027$ & $0.0384 \pm 0.0003$ & $0.03829 \pm 0.00046$ \\  [2pt]%
Inclination \dotfill & $i$ & degree & $89.10 \pm 0.24$ & $89.0 \pm 0.4$ & $89.915 \pm 0.098$ \\ [2pt] %
RV-curve semi-amplitude \dotfill & $K_{\rm A}$ &  m\,s$^{-1}$ & $-$ & $35.8 \pm 1.9$ & -- \\ [2pt] %
Barycentric RV  \dotfill & $\gamma$ & km\,s$^{-1}$ & $-40.4589 \pm 0.0023  $   & $-40.51 \pm 0.21$  & --\\  [2pt] %
Projected spin-orbit angle  \dotfill & $\lambda$ & degree  & $-54^{+41}_{-13}$ & -- & -- \\ [2pt] %
\hline %
\end{tabular}
\end{table*}

\begin{table*}
\tiny
\centering %
\caption{Physical parameters of the planetary system HAT-P-22 derived in this work.} %
\label{tab:hatp22_final_parameters} %
\begin{tabular}{l c c c c c}
\hline %
\hline  \\[-8pt]
Parameter & Nomen. & Unit & This Work & \citet{bakos:2011} & \citet{turner:2016} \\
\hline  \\[-6pt]
\multicolumn{1}{l}{\textbf{Stellar parameters}} \\[2pt] %
Effective temperature         \dotfill & $T_{\mathrm{eff}}$   & K            & $5314 \pm 50$ & $5302 \pm 80$ & -- \\ [2pt] %
Iron abundance               \dotfill & [Fe/H]               &              & $+0.30 \pm 0.09$ & $+0.24 \pm 0.08$ & -- \\ [2pt] %
Projected rotational velocity \dotfill & $v\,\sin{i_{\star}}$ & km\,s$^{-1}$ & $1.3 \pm 0.7 $ & $0.5 \pm 0.5$  & -- \\ [2pt] %
Mass    \dotfill & $M_{\star}$ & $M_{\sun}$ & $0.936 \pm 0.028 \pm 0.033$ & $0.916 \pm 0.035 $ & -- \\ [2pt]%
Radius  \dotfill & $R_{\star}$ & $R_{\sun}$ & $1.062 \pm 0.046 \pm 0.013$ & $1.040 \pm 0.044$ & -- \\ [2pt] %
Mean density \dotfill & $\rho_{\star}$ & $\rho_{\sun}$ & $0.781 \pm 0.099$ & -- & --\\ [2pt] %
Logarithmic surface gravity \dotfill & $\log{g_{\star}}$ & cgs & $4.357 \pm 0.039 \pm 0.005$ & $4.36 \pm 0.04$ & -- \\ %
Age \dotfill & & Gyr & $9.0_{-2.2\,-3.0}^{+1.4\,+3.7} $ & $12.4 \pm 2.6$\\ [2pt]  %
\hline \\[-6pt]%
\multicolumn{1}{l}{\textbf{Planetary parameters}} \\ [2pt] %
Mass \dotfill   & $M_{\mathrm{p}}$ & $M_{\mathrm{Jup}}$ & $2.192 \pm 0.057 \pm 0.052$ & $2.147 \pm 0.061$ & $2.148 \pm 0.062$ \\ [2pt] %
Radius \dotfill & $R_{\mathrm{p}}$ & $R_{\mathrm{Jup}}$ & $1.060 \pm 0.073 \pm 0.013$ & $1.080 \pm 0.058 $ & $1.092 \pm 0.047 $ \\ [2pt] %
Mean density \dotfill & $\rho_{\mathrm{p}}$ & $\rho_{\mathrm{Jup}}$ & $1.72 \pm 0.35 \pm 0.02$ & $1.59^{+0.3}_{-0.22}$&$1.61\pm0.21$ \\[2pt]%
Surface gravity  \dotfill & $g_{\mathrm{p}}$ & m\,s$^{-2}$ & $48.3 \pm  6.6$ & $45.7 \pm 5.3$ & $49^{+8}_{-7}$ \\ [2pt] %
Equilibrium temperature \dotfill & $T_{\mathrm{eq}}$ & K & $1293 \pm 29$ & $1283 \pm 32$ & -- \\  [2pt] %
Safronov number \dotfill & $\Theta$ & & $0.184 \pm 0.013 \pm 0.002$ & $0.179 \pm 0.010$ & -- \\  [2pt]%
\hline \\[-6pt]%
\multicolumn{1}{l}{\textbf{Orbital parameters}} \\ [2pt] %
Time of mid-transit \dotfill & $T_{0}$ & BJD(TDB)& $2\,454\,930.22016\,(16)$ & $2\,454\,931.809\,(16)$ & 2\,454\,930.22296\,(25) \\ [2pt] %
Period              \dotfill & $P_{\mathrm{orb}}$ & days & $3.21223328\,(58)$ & $3.212220\,(9)$ & 3.2122312\,(12) \\  [2pt]%
Semi-major axis \dotfill & $a$ & au & $0.04171 \pm 0.00042 \pm 0.00050$ & $0.0414 \pm 0.0005$ & -- \\  [2pt]%
Inclination \dotfill & $i$ & degree & $86.46 \pm 0.41$ & $86.9^{+0.6}_{-0.5}$ & -- \\ [2pt] %
RV-curve semi-amplitude \dotfill & $K_{\rm A}$ &  m\,s$^{-1}$ & -- & $313.3 \pm 4.2$ & -- \\ [2pt] %
Barycentric RV  \dotfill & $\gamma$ & km\,s$^{-1}$ & $+12.63696 \pm 0.00035  $   & $+12.49 \pm 0.28$  & --\\  [2pt] %
Projected spin-orbit angle  \dotfill & $\lambda$ & degree  & $-2.1 \pm 3.0$ & -- & -- \\ [2pt] 
True spin-orbit angle  \dotfill & $\psi$      & degree  & $1.5^{\circ}\,^{+30.0^{\circ}}_{-1.5^{\circ}}$ & -- & -- \\  [2pt] %
\hline %
\end{tabular}
\end{table*}

\begin{table*}
\tiny
\centering %
\caption{Physical parameters of the planetary system WASP-39 derived in this work.} %
\label{tab:wasp39_final_parameters} %
\begin{tabular}{l c c c c c}
\hline %
\hline  \\[-8pt]
Parameter & Nomen. & Unit & This Work & \citet{faedi:2011} & \citet{maciejewski:2016} \\
\hline  \\[-6pt]
\multicolumn{1}{l}{\textbf{Stellar parameters}} \\[2pt] %
Effective temperature         \dotfill & $T_{\mathrm{eff}}$   & K & $5485 \pm 50$   & $5400  \pm 150$  & -- \\ [2pt] %
Iron abundance               \dotfill & [Fe/H]               &   & $+0.01 \pm 0.09$ & $-0.12 \pm 0.10$ & -- \\ [2pt] %
Projected rotational velocity \dotfill & $v\,\sin{i_{\star}}$ & km\,s$^{-1}$ & $1.0 \pm 0.5$ & $1.4 \pm 0.6$  & -- \\ [2pt] %
Mass    \dotfill & $M_{\star}$ & $M_{\sun}$ & $0.913 \pm 0.035 \pm 0.031$ & $0.93 \pm 0.03$ & -- \\ [2pt]%
Radius  \dotfill & $R_{\star}$ & $R_{\sun}$ & $0.939 \pm 0.019 \pm 0.011$ & $0.895 \pm 0.023$ & $0.918^{+0.022}_{-0.019}$ \\ [2pt] %
Mean density \dotfill & $\rho_{\star}$ & $\rho_{\sun}$ & $1.103 \pm 0.057$ & $1.297^{+0.082}_{-0.074}$ & $1.201^{+0.075}_{-0.063}$ \\ [2pt] %
Logarithmic surface gravity \dotfill & $\log{g_{\star}}$ & cgs & $4.453 \pm 0.017 \pm 0.005$&$4.503\pm0.017$&$4.480^{+0.029}_{-0.025}$ \\ %
Age \dotfill & & Gyr & $8.5_{-1.0\,-3.3}^{+3.5\,+2.0} $ & $9^{+3}_{-4}$ & --\\ [2pt]  %
\hline \\[-6pt]%
\multicolumn{1}{l}{\textbf{Planetary parameters}} \\ [2pt] %
Mass \dotfill   & $M_{\mathrm{p}}$ & $M_{\mathrm{Jup}}$ & $0.281 \pm 0.031 \pm 0.006$ & $0.28 \pm 0.03$ & $0.283 \pm 0.041$ \\ [2pt] %
Radius \dotfill & $R_{\mathrm{p}}$ & $R_{\mathrm{Jup}}$ & $1.279 \pm 0.037 \pm 0.014 $ & $1.27 \pm 0.04$&$1.332^{+0.034}_{-0.031}$ \\ [2pt] %
Mean density \dotfill & $\rho_{\mathrm{p}}$ & $\rho_{\mathrm{Jup}}$ &$0.126\pm 0.017\pm0.001$&$0.14\pm0.02$&$0.120^{+0.020}_{-0.019}$ \\[2pt]%
Surface gravity  \dotfill & $g_{\mathrm{p}}$ & m\,s$^{-2}$ & $4.26 \pm 0.50$ & $4.07 \pm 0.46$ & $4.14^{+0.62}_{-0.61}$ \\ [2pt] %
Equilibrium temperature \dotfill & $T_{\mathrm{eq}}$ & K & $1166 \pm 14$ & $1116^{+33}_{-32}$ & -- \\  [2pt] %
Safronov number \dotfill & $\Theta$ & & $0.0232 \pm 0.0025 \pm 0.0003$ & \textendash & -- \\  [2pt]%
\hline \\[-6pt]%
\multicolumn{1}{l}{\textbf{Orbital parameters}} \\ [2pt] %
Time of mid-transit \dotfill & $T_{0}$ & BJD(TDB)& $2\,455\,342.96913\,(63)$ & $2\,455\,342.9688\,(2)$ & 2\,455\,342.96982\,(51) \\ [2pt] %
Period              \dotfill & $P_{\mathrm{orb}}$ & days & $4.0552941\,(34)$ & $4.055259\,(9) $ & $4.0552765\,(35)$ \\  [2pt]%
Semi-major axis \dotfill & $a$ & au & $0.04828 \pm 0.00061 \pm 0.00054$ & $0.0486 \pm 0.0005$ & $0.04858 \pm 0.00052$ \\  [2pt]%
Inclination \dotfill & $i$ & degree & $87.32 \pm 0.17$ & $87.83^{+0.25}_{-0.22}$ & $87.75^{+0.27}_{-0.20}$ \\ [2pt] %
RV-curve semi-amplitude \dotfill & $K_{\rm A}$ &  m\,s$^{-1}$ & -- & $38 \pm 4$ & $37.9 \pm 5.4$ \\ [2pt] %
Barycentric RV  \dotfill & $\gamma$ & km\,s$^{-1}$ & $-58.4421 \pm 0.0020$ & $-58.4826 \pm 0.0004$  & --\\  [2pt] %
Projected spin-orbit angle  \dotfill & $\lambda$ & degree  & $0 \pm 11$ & -- & -- \\ [2pt] 
\hline %
\end{tabular}
\end{table*}

\begin{table*}
\tiny
\centering %
\caption{Physical parameters of the planetary system WASP-60 derived in this work.} %
\label{tab:wasp60_final_parameters} %
\begin{tabular}{l c c c c c}
\hline %
\hline  \\[-8pt]
Parameter & Nomen. & Unit & This Work & \citet{hebrard:2013} & \citet{turner:2017}  \\
\hline  \\[-6pt]
\multicolumn{1}{l}{\textbf{Stellar parameters}} \\[2pt] %
Effective temperature         \dotfill & $T_{\mathrm{eff}}$   & K & $6105 \pm 50$   & $5900  \pm 100$ & --  \\ [2pt] %
Iron abundance               \dotfill & [Fe/H]               &   & $+0.26 \pm 0.07$ & $-0.04 \pm 0.09$ & -- \\ [2pt] %
Projected rotational velocity \dotfill & $v\,\sin{i_{\star}}$ & km\,s$^{-1}$ & $3.8 \pm 0.6$ & $3.4 \pm 0.8$ & --\\ [2pt] %
Mass    \dotfill & $M_{\star}$ & $M_{\sun}$ & $1.229 \pm 0.026 \pm 0.015$ & $1.078 \pm 0.035$ & -- \\ [2pt]%
Radius  \dotfill & $R_{\star}$ & $R_{\sun}$ & $1.401 \pm 0.066 \pm 0.006$ & $1.14 \pm 0.13 $ & -- \\ [2pt] %
Mean density \dotfill & $\rho_{\star}$ & $\rho_{\sun}$ & $0.447 \pm 0.063$ & $0.72 \pm 0.20$ & --\\ [2pt] %
Logarithmic surface gravity \dotfill & $\log{g_{\star}}$ & cgs & $4.235 \pm 0.041 \pm 0.002$ & $4.35 \pm 0.09$ & -- \\ %
Age \dotfill & & Gyr & $1.7_{-0.5\,-0.2}^{+0.5\,+0.4} $ & $9^{+3}_{-4}$ & -- \\ [2pt]  %
\hline \\[-6pt]%
\multicolumn{1}{l}{\textbf{Planetary parameters}} \\ [2pt] %
Mass \dotfill   & $M_{\mathrm{p}}$ & $M_{\mathrm{Jup}}$ & $0.560 \pm 0.036 \pm 0.005$ & $0.514 \pm 0.034$ & $0.512 \pm 0.034$ \\ [2pt] %
Radius \dotfill & $R_{\mathrm{p}}$ & $R_{\mathrm{Jup}}$ & $1.225 \pm 0.069 \pm 0.005$ & $0.86  \pm 0.12$ & $ 0.94 \pm 0.12$ \\ [2pt] %
Mean density \dotfill & $\rho_{\mathrm{p}}$ & $\rho_{\mathrm{Jup}}$ & $0.285 \pm 0.052 \pm 0.001$ & $0.8 \pm 0.3$ & $0.75\pm0.27$ \\[2pt]%
Surface gravity  \dotfill & $g_{\mathrm{p}}$ & m\,s$^{-2}$ & $9.2 \pm 1.2$ & $15.5^{+4.9}_{-3.7}$ & $12.8 \pm 6.5$\\ [2pt] %
Equilibrium temperature \dotfill & $T_{\mathrm{eq}}$ & K & $1479 \pm 35$ & $1320 \pm 75$ &  $1354\pm23$ \\  [2pt] %
Safronov number \dotfill & $\Theta$ & & $0.0411 \pm 0.0036 \pm 0.0002$ & \textendash & $0.051 \pm 0.013$ \\  [2pt]%
\hline \\[-6pt]%
\multicolumn{1}{l}{\textbf{Orbital parameters}} \\ [2pt] %
Time of mid-transit \dotfill & $T_{0}$ & BJD(TDB)& $2\,456\,952.43264\,(17)$ & $2\,455\,747.0295\,(22)$ & $2\,455\,747.0302\,(22)$ \\ [2pt] %
Period              \dotfill & $P_{\mathrm{orb}}$ & days & $4.3050040\,(59)$ & $4.3050011\,(62)$ & 4.305022\,(21) \\  [2pt]%
Semi-major axis \dotfill & $a$ & au & $0.05548 \pm 0.00040  \pm 0.00023$ & $0.0531 \pm 0.0006$ & $0.050 \pm 0.011$ \\  [2pt]%
Inclination \dotfill & $i$ & degree & $86.10 \pm 0.61$ & $87.9 \pm 1.6$ & $87.48 \pm 2.83$ \\ [2pt] %
RV-curve semi-amplitude \dotfill & $K_{\rm A}$ &  m\,s$^{-1}$ & -- & $60.8 \pm 3.8$ & --\\ [2pt] %
Barycentric RV  \dotfill & $\gamma$ & km\,s$^{-1}$ & $-26.532 \pm 0.021$ & -- & --\\  [2pt] %
Projected spin-orbit angle  \dotfill & $\lambda$ & degree  & $-129 \pm 17$ & -- & --\\ [2pt] 
\hline %
\end{tabular}
\end{table*}

\end{appendix}

\end{document}